\documentclass[aps,prc,twocolumn,floatfix,nofootinbib,showpacs,showkeys,10pt]{revtex4-1}

\usepackage{amsmath}
\usepackage{graphicx}
\usepackage{float}
\usepackage[colorlinks = true,linkcolor = blue,urlcolor  = blue,citecolor = blue,anchorcolor = blue]{hyperref}
\usepackage{dcolumn}

\begin{document}

\title{Effects of the equation of state on the bulk properties of maximally-rotating neutron stars}

\author{P.S. Koliogiannis}
\email{pkoliogi@physics.auth.gr}

\author{Ch.C. Moustakidis}
\email{moustaki@auth.gr}

\affiliation{Department of Theoretical Physics, Aristotle University of Thessaloniki, 54124 Thessaloniki, Greece}
\begin{abstract}
Neutron stars are among the densest known objects in the universe and an ideal laboratory for the strange physics of super-condensed matter. While the simultaneously measurements of mass and radius of non-rotating neutron stars may impose constraints on the properties of the dense nuclear matter, the observation and study of maximally-rotating ones, close to the mass-shedding limit, may lead to significantly further constraints. Theoretical predictions allow neutron stars to rotate extremely fast (even more than $2000 \ {\rm Hz}$). However, until this moment, the fastest observed rotating pulsar has a frequency of $716 \ {\rm Hz}$, much lower compared to the theoretical predictions. There are many suggestions for the mechanism which lead to this situation. In any case, the theoretical study of uniformly rotating neutron stars, along with the accurate measurements, may offer rich information concerning the high density part of the equation of state. In addition, neutron stars through their evolution, may provide us with a criteria to determine the final fate of a rotating compact star. Sensitivity of bulk neutron stars properties on the equation of state at the mass-shedding limit are the main subject of the present study.

\pacs{26.60.-c, 26.60.Kp, 21.65.Mn, 97.60.Gb}

\keywords{Nuclear equation of state; Neutron stars; Uniform rotation; Keplerian sequence}
\end{abstract}

\maketitle

\section{Introduction}\label{sec:1}
Neutron stars are considered as extraordinary astronomical laboratories for the physics of nuclear matter because, from an astrophysical point of view, are the objects with the most fascinating constitution of energy and matter in the Universe~\cite{Shapiro-83,Glendenning-2000,Haensel-07}. To be more specific, the observation of mass, as well as the radius, of slowly-rotating (or non-rotating) neutron stars may provide us with useful constraints on the equation of state (EoS) of nuclear matter. In addition, neutron stars, due to their compactness, may rotate very fast compared to other astrophysical objects~\cite{Friedman-13}. Henceforth, measurements of specific properties (including mainly the mass and radius, frequency, moment of inertia, quadrupole moment etc.) of maximally-rotating neutron stars, close to mass-shedding limit (hereafter maximally-rotating corresponds to the configuration close or at the Keplerian frequency) may lead to robust constraints on the EoS as well as on the constitution of nuclear matter at high densities.\\
\indent The determination of the maximum neutron star mass is a long-standing issue in astrophysics since it is directly related with the identification of black holes and the unknown behavior of the nuclear matter at high densities. Until this moment, the most massive neutron stars measurements (in solar masses, $M_{\odot}$) include: a) the PSR J1614-2230 ($M=1.97\pm 0.04 \ M_{\odot}$)~\cite{Demorest-010} (or from recent elaboration of the observation $M=1.928\pm 0.017\ M_{\odot}$~\cite{Fonseca-016} and also $M=1.908\pm 0.016 M_{\odot}$~\cite{Arzoumanian-2018}), b) the PSR J0348+0432 ($M=2.01\pm 0.04 \ M_{\odot}$)~\cite{Antoniadis-013}, c) the PSR J0740+6620 ($M=2.14^{+0.10}_{- 0.09} \ M_{\odot}$)~\cite{Cromartie-19} and d) the PSR J2215+5135 ($M=2.27^{+0.17}_{- 0.15} \ M_{\odot}$)~\cite{Linares-18}. In addition, there is a detailed study concerning the spin frequency of rotating neutron stars~(for  a review see Refs.~\cite{Patruno-2017}). The fastest rotating pulsar that has been found is the J1748-244ad with a spin frequency of $716 \ {\rm Hz}$~\cite{Hessels-06}. However, the issue is still open: {\it why we have not observed pulsars with higher values of frequency which predicted from the majority of theoretical models?} And even more, {\it what limits the spin frequencies of millisecond pulsars and why?}~\cite{Prakash-015}. Future measurements of moment of inertia~\cite{Beiger-2005} and Keplerian frequency may be the answer to these questions by improving considerably our knowledge on the properties of maximally-rotating neutron stars.\\
\indent The effects of the EoS on the properties of rotating neutron stars (see Refs.~\cite{Stergioulas-1998,Paschalidis-2017} for introduction and relevant bibliography) had begun to gain ground almost thirty years ago from Shapiro, Teukolsky and their colleagues~\cite{Shapiro-89,Cook-92,Cook-94a,Cook-94b,Cook-94c}. A significant contribution on these issues had been also made from Friedman and his colleagues~\cite{Koranda-97,Friedman-87,Friedman-86,Friedman-88,Friedman-89}, Haensel and co-workers~\cite{Salgado-94a,Salgado-94b,Haensel-89,Haensel-95,Lasota-96,Haensel-99}, as well as Glendenning and his colleagues~\cite{Weber-91,Weber-92,Glendening-92,Glendening-94}. Rapid rotation and its effects on the EoS had been studied also in Refs.~\cite{Wagover-78,Lindblom-86,Lattimer-1990,Hashimoto-94,Iida-1997,Shibata-2000} and most recently in Refs.~\cite{Benhar-05,Dhiman-07,Agrawal-08,Krastev-2008,Haensel-08,Haensel-09,Lo-2011,Zhang-013,Chakrabarti-2014,
Cipolletta-015,Breu-016,Haensel-016,Bejger-017,Cipolletta-2017,Haskell-018,Riahi-019}. Moreover, in nuclear astrophysics hot neutron stars in correlation with rapid rotation had been studied in Refs.~\cite{Marques-2017,Batra-2018}. In addition, maximally-rotating neutron stars in modified gravity theories have been studied in detail by Kokkotas and his colleagues~\cite{Doneva-2013,Yazadjiev-2015}.\\
\indent In this work we extend the previous fundamental work of Cook, Shapiro and Teukolsky~\cite{Cook-94b}, as well as the most recent work of Cipolletta {\it et al.}~\cite{Cipolletta-015}. In particular, we employ a large number of modern EoSs (combined with a few previous ones) which  all of them, at least marginally (few of them), predict the upper bound of the maximum neutron star mass of $M=1.908\pm 0.016\ M_{\odot}$~\cite{Arzoumanian-2018}, while also reproducing accurately the bulk properties of symmetric nuclear matter (for more details see Ref.~\cite{Koliogiannis-19}). The models of these EoSs are phenomenological, field theoretical and microscopic. In the category of phenomenological models, there are the: MDI~\cite{Prakash-1997,Moustakidis-08}, HHJ~\cite{Heiselberg -2000}, Ska, SkI4~\cite{Chabanat-97,Farine-97} and DH~\cite{Douchin-01}, in field theoretical one, there are the: NLD~\cite{Gaitanos-013,Gaitanos-015} and W~\cite{Walecka-74} and in microscopic one, there are the: HLPS (based on nuclear interactions derived from chiral effective field theory)~\cite{Hebeler-013}, SCVBB (using the Argonne v18 potential plus three-body forces computed with the Urbana model)~\cite{Sharma-015}, BS~\cite{Balberg-2000}, BGP (Relativistic pion exchange)~\cite{Bowers-75}, BL~\cite{Bombaci-018}, WFF1,WFF2~\cite{Wiringa-88} and PS~\cite{Pandha-75}. It has to be stressed out that the majority of the mentioned EoSs have been constructed in order to reproduce the bulk properties of uniform symmetric nuclear matter and also to extend to pure neutron matter. The extension to neutron star matter is performed with respect to beta equilibrium. As far as concerning the leptonic degree of freedom, in most of them it is considered that the main contribution of leptons is due to electrons. All of the used EoSs are properly describe the fluid core of a neutron star. It should be noted also that few of them have been applied firstly for the study of finite nuclei. Among the number of equations that we use, we have construct two EoSs, the APR-1 and APR-2 (Microscopic model)~\cite{Akmal-98}, predicted by the Momentum-Dependent Interaction model (MDI). This model reproduces the results of microscopic calculations of symmetric nuclear matter and neutron star matter at zero temperature with the advantage of its extension to finite temperature. For the solid crust region of all the EoSs we employed the EoS of Feynman, Metropolis and Teller~\cite{Feynman-1949} and also of Baym, Bethe and Sutherland ~\cite{Baym-1971}.\\
\indent An effort was made to systematically study the most of the bulk properties of uniformly rotating neutron stars at the Keplerian sequence (the sequence in which the maximum mass configuration corresponds to the Keplerian frequency), including the mass, polar and equatorial radius, angular velocity, moment of inertia, Kerr parameter, eccentricity, braking index and etc. Additionally, for reasons of completeness and comparison, because all EoSs that we use are hadronic ones, we present also an EoS with appearence of hyperons at high densities (FSU2H)~\cite{Tolos-2017} and one suitable to describe quark stars based on MIT bag model (QS57.6)~\cite{Glendenning-2000,Haensel-07}.\\
\indent Furthermore, we explore the possibility to update the previous empirical universal relations which connecting  the Keplerian frequency with the mass and radius at the maximum mass configuration. We systematically study the Kerr parameter dependence on the EoS and also provide the evolution of the angular momentum of a neutron star in order to examine the case where neutron stars considered to be progenitors of black holes. In particular, we examine (according to the terminology of Ref.~\cite{Cook-94b}) two equilibrium sequences of rotating neutron stars, {\it normal} and {\it  supramassive}. While {\it normal} evolutionary sequences have a spherical, non-rotating (stable) end point, {\it  supramassive} ones, which by definition have masses higher than the maximum mass of the non-rotating neutron star, they don't have a stable end point and as a consequence, the collapse to a black hole is inevitable. However, the construction of {\it normal} and mainly {\it  supramassive} sequences is a complicated procedure in the framework of General Relativity~\cite{Cook-94b}.\\
\indent In addition, we systematically study the moment of inertia, a quantity which plays important role on the properties of rotating neutron stars, and eccentricity which can inform us for their deformation. Following the previous work of Lattimer and Prakash~\cite{Lattimer-05}, we also provide an absolute upper limit of the higher density of cold baryonic matter in the Universe, based on the upper limit imposed by the maximum mass of a neutron star. In fact, we try to improve the bound which was introduced in Ref.~\cite{Lattimer-05}, by using updated EoSs and including also the case of maximally-rotating neutron stars. Finally, we study the effects of the EoS on the braking index of pulsars. We mainly focus on values near the Keplerian frequency (70\% and more) where the braking index begins to be affected by the rest mass (definition has been given in a proper section).\\
\indent The article is organized as follows. In Section \ref{sec:2} we briefly review the properties of nuclear matter, the computational hypothesis and the models for the nuclear EoSs. In Section \ref{sec:3} we present the rotating configuration for neutron stars. In particular, we introduce the effects of the Keplerian frequency on the bulk properties of neutron stars and we also describe two properties of the EoS, moment of inertia and eccentricity. In addition, we provide a discussion for the Kerr parameter and the fully described rest mass sequences. The upper bound for density of cold baryonic matter and the effects of the braking index on the EoS are also obtained. Section \ref{sec:4} contains the discussion and main conclusions of the present study. Finally, useful expressions and clarifications are given in Appendix.

\section{The nuclear equation of state}\label{sec:2}
In the present study we have suitably selected and employed a large number of hadronic EoSs~\cite{Koliogiannis-19}. Moreover, we have constructed two additional EoSs, the APR-1 and APR-2, by using the MDI model (for more details see Appendix~\ref{app:1}) and data from Akmal {\it et al.}~\cite{Akmal-98}. Except the hadronic EoSs, an EoS with appearance of hyperons at high densities and one suitable to describe quark stars have been used for completeness~\cite{Glendenning-2000,Haensel-07,Tolos-2017}.\\
\indent There are many reasons to support the reliability of the MDI model. In particular,
a) reproduces with high accuracy the properties of symmetric nuclear matter at the saturation density, b) the theoretical prediction of the value and slope of symmetry energy at the saturation density are close to the experimental predictions, c) reproduces other properties of symmetric nuclear matter (SNM) (including isovector quantities $K_{0}$ and $Q_{0}$) inside the limiting area of the experimental data, d) reproduces correctly the microscopic calculation of the Chiral model~\cite{Hebeler-10} for pure neutron matter (PNM) (for low densities) and the results of \emph{state-of-the-art} calculations of Akmal {\it et al.}~\cite{Akmal-98} (for high densities), e) has the flexibility that the energy per particle depends not only on the density, but also on the momentum, f) can be easily extended to include temperature dependence (which is needed to study core-collapse supernova, proto-neutron stars, neutron stars merger etc.) and g) predicts maximum neutron star mass higher than the observed ones~\cite{Demorest-010,Fonseca-016,Arzoumanian-2018,Antoniadis-013,Cromartie-19,Linares-18}.

\subsection{Properties of nuclear matter and the construction of the APR-1 and APR-2 EoSs}\label{secsub:2_1}
\indent Assuming that the neutron-proton asymmetry is characterized by the parameter~\cite{Constantinou-2014,Constantinou-2015}
\begin{equation}
I = \frac{n_{n}-n_{p}}{n} = 1-2x
\end{equation}
where $n_{p}$, $n_{n}$ and $n=n_{p}+n_{n}$ are the proton, neutron and total densities and $x = n_{p}/n$ is the proton fraction, the total energy per particle can be expanded as follows
\begin{equation}
E(n,I) = E(n,0)+\sum_{k=2,4,\cdots}E_{{\rm sym},k}(n)I^{k}
\label{Expand-1}
\end{equation}
where
\begin{equation}
E_{{\rm sym},k}(n)=\left. \frac{1}{k!}\frac{\partial^{k} E(n,I)}{\partial I^{k}}\right|_{I=0}
\label{Esym-def}
\end{equation}
\indent We studied two cases in this paper, the parabolic (symbolized as {\it pa}) and the full (symbolized as {\it f}) approximation. In the case of the parabolic approximation we considered that the symmetry energy is given through
\begin{equation}
E_{\rm sym, pa}(n) = E(n,I = 1) - E(n,I = 0)
\label{PA-1}
\end{equation}
while in the full approximation it is given through
\begin{equation}
E_{\rm sym,f}(n) = E_{{\rm sym},2}(n) = S_{2}(n)
\label{FULL-1}
\end{equation}

\indent The properties of nuclear matter at the saturation density are defined as~\cite{Constantinou-2014,Constantinou-2015}

\begin{equation}
L = 3n_s\left. \frac{dS_{2}(n)}{dn}\right|_{n_s}, \quad K = \left.9n_{s}^{2} \frac{d^{2} S_{2} (n)}{dn^{2}}\right|_{n_{s}},
\label{K-1}
\end{equation}

\begin{equation}
Q = 27n_s^3\left. \frac{d^{3} S_{2} (n)}{dn^{3}}\right|_{n_{s}}, \quad  K_{0} = \left.9n_s^2\frac{d^2 E(n,0)}{dn^2}\right|_{n_s},
\label{L-1}
\end{equation}
\begin{equation}
Q_{0} = 27n_{s}^{3}\left. \frac{d^{3} E(n,0)}{dn^{3}}\right|_{n_{s}},
\label{Qsym-1}
\end{equation}
where $L$, $K$, $Q$ are related to the first, second and third derivative of the symmetry energy $S_{2}(n)$, respectively. $K_{0}$ is the compression modulus and $Q_{0}$ is related to the third derivative of $E(n,0)$. The $n_{s}$ is the saturation density of symmetric nuclear matter and its equal to 0.16 $\rm fm^{-3}$. The last property is the ratio of the Landau effective mass to mass in vacuum for the MDI model~\cite{Prakash-1997,Moustakidis-15,Constantinou-2014,Constantinou-2015,Moustakidis-2008}, and given by
\begin{widetext}
	\begin{equation}
	\frac{m^{*}_{\tau}(n,I)}{m_{\tau}} = \left[ 1 - \frac{2nm_{\tau}}{n_{s}\hbar^{2}} \sum_{i=1,2} \frac{1}{\Lambda_{i}^{2}}\frac{C_{i} \pm \frac{C_{i}-8Z_{i}}{5}I}{\left[1 + \left(\frac{k_{F}^{0}}{\Lambda_{i}}\right)^{2} \left[\left(1 \pm I\right)\frac{n}{n_{s}}\right]^{2/3}\right]^{2}}\right]^{-1}
	\label{eq:lem}
	\end{equation}
\end{widetext}
where $\tau$ corresponds to neutrons or protons.\\
\indent Although we studied two cases, the parabolic and full approximation, the one that is used in the detailed study is the full approximation. Both of them lead to similar results. However, the parabolic approximation is referenced for future studies.\\
\indent The parametrization of the MDI model (Eq.~\eqref{e-T0}) is performed by using data originated from previous work of Akmal {\it et al.}~\cite{Akmal-98}. In particular, we employed the data concerning the energy per particle of symmetric and pure neutron matter (in the area $0.04 \ {\rm fm}^{-3} \leq n \leq 0.96 \ {\rm fm}^{-3}$) and for models ${\rm A}18+{\rm UIX}$ (hereafter APR-1)  and ${\rm A}18+\delta v+{\rm UIX}^*$ (hereafter APR-2). In order to achieve the best fitting to Akmal's data using the Eq.~\eqref{e-T0}, we divided our region of study in three sections : a) Low Density Region ($0.04 \ {\rm fm}^{-3} \leq n \leq 0.2 \ {\rm fm}^{-3}$), b) Medium Density  Region ($0.2 \ {\rm fm}^{-3} \leq n \leq 0.56 \ {\rm fm}^{-3}$) and c) High Density Region ($0.56 \ {\rm fm}^{-3} \leq n \leq 0.96 \ {\rm fm}^{-3}$). With this method we have calculated the coupling constants and the parameters for the asymmetric nuclear matter.\\
\indent The main properties of nuclear matter at the saturation density $n_s$, calculated with Eq.~\eqref{PA-1} through Eq.~\eqref{eq:lem} for the APR-1 and APR-2 EoSs, including also isovector quantities, are presented in Table \ref{tab:1}. It should be noted that the parametrization of pure neutron matter leads to results in agreement with the predictions of chiral effective field theory~\cite{Hebeler-10} for low densities. For high density values, the parametrization leads to the prediction of Akmal {\it et al.}~\cite{Akmal-98}. The main drawback of these two EoSs is related with the violation of causality; the speed of sound becomes greater than the speed of light at high densities. However, the parametrization of the MDI model, has the advantage that prevents the onset from violate the causality.

\begin{table}[H]
\squeezetable
\caption{Properties of nuclear matter (NM) for APR-1 and APR-2 EoSs.}
\begin{ruledtabular}
\begin{tabular}{ccc}
Properties of NM & APR-1 & APR-2\\
\hline
$L_{pa}$ (MeV) & 63.18 & 57.43\\

$Q_{pa}$ (MeV) & 482.34 & 568.91\\

$K_{pa}$ (MeV) & -103.70 & -118.78\\

$E_{sym_{pa}}$ (MeV) & 33.61 & 33.59\\

$L_{f}$ (MeV) & 63.31 & 57.40\\

$Q_{f}$ (MeV) & 450.50 & 538.44\\

$K_{f}$ (MeV) & -88.26 & -99.81\\

$E_{sym_{f}}$ (MeV) & 32.74 & 32.53\\

$Q_{0}$ (MeV) & -581.27 & -581.27\\

$K_{0}$ (MeV) & 256.40 & 256.40\\

$m^{\ast}_{\tau}/m_{\tau}$ & 0.72 & 0.72
\end{tabular}
\end{ruledtabular}
\label{tab:1}
\end{table}

\begin{figure}[H]
	\includegraphics[width=0.5\textwidth]{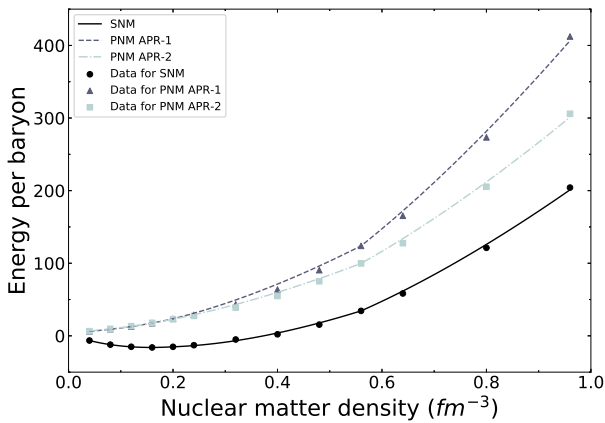}
	\caption{(color online) Symmetric nuclear matter and pure neutron matter fits for APR EoSs using Akmal's~\cite{Akmal-98} data and the MDI model. The SNM is presented with the circles and the solid line, the APR-1 PNM is presented with the triangles and the dashed line and the APR-2 PNM is presented with the squares and the dashed-dotted line.}
	\label{fig:nm}
\end{figure}

The schematic presentation of Eq.~\eqref{e-T0} for APR EoSs and the data from Akmal {\it et al.}~\cite{Akmal-98} are presented in Fig.~\ref{fig:nm}.

\subsection{The selected Equations of State}\label{secsub:2_2}
\begin{figure}[H]
\includegraphics[width=0.5\textwidth]{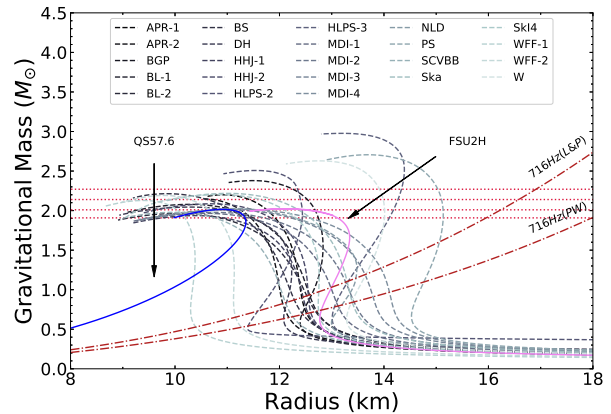}
\caption{(color online) Mass - Radius diagram for the 23 EoSs at the non-rotating configuration. The observed limits of neutron star mass are presented with the horizontal dotted lines ($1.908 M_{\odot}$, $2.01 M_{\odot}$, $2.14 M_{\odot}$ and $2.27 M_{\odot}$). The observed limit of 716 Hz, from Lattimer and Prakash (L\&P)~\cite{Lattimer-2004} and from the present work (PW) (for more details see Appendix \ref{app:2}), is presented with the curved dashed-dotted lines. The two indicated solid lines correspond to the EoS with appearance of hyperons at high densities (FSU2H) and the one suitable to describe quark stars (QS57.6).}
\label{fig:mass_radius}
\end{figure}

The EoSs that we used~\cite{Koliogiannis-19,Prakash-1997,Moustakidis-08,Heiselberg -2000,Chabanat-97,Farine-97,Douchin-01,Gaitanos-013,Gaitanos-015,Walecka-74,Hebeler-013,Sharma-015,Balberg-2000,Bowers-75,Bombaci-018,Wiringa-88,Pandha-75,Akmal-98} are in consistent with the current observed limits of neutron star mass~\cite{Demorest-010,Fonseca-016,Arzoumanian-2018,Antoniadis-013,Cromartie-19,Linares-18} and also with the one for frequency~\cite{Hessels-06}. In Fig. \ref{fig:mass_radius} we present the gravitational mass versus the corresponding equatorial radius (hereafter radius) for the 23  EoSs at the non-rotating configuration, where the current observed limits are also presented. Moreover, the EoS with appearance of hyperons at high densities (FSU2H) and the one suitable to describe quark stars (QS57.6) are also indicated.

\section{Rotating Neutron Stars}\label{sec:3}
In the framework of General Relativity rotating neutron stars can be described a) by the stationary axisymmetric space-time metric~\cite{Friedman-13,Paschalidis-2017}
\begin{equation}
ds^{2} = -e^{2\nu} dt^{2} + e^{2\psi}\left(d\phi - \omega dt\right)^{2} + e^{2\mu} \left(dr^{2} + r^{2}d\theta^{2}\right)
\end{equation}
where the metric functions $\nu$, $\psi$, $\omega$ and $\mu$ depend only on the coordinates $r$ and $\theta$, and b) the matter inside the neutron star. If we neglect sources of non-isotropic stresses, as well as viscous ones and heat transport, then the matter inside the neutron star can be fully described by the stress-energy tensor and modeled as a perfect fluid~\cite{Friedman-13,Paschalidis-2017},
\begin{equation}
T^{\alpha \beta} = \left(\varepsilon + P\right) u^{\alpha} u^{\beta} + P g^{\alpha \beta}
\end{equation}
where $u^{\alpha}$ is the fluid's 4-velocity. The energy density and pressure is denoted as $\varepsilon$ and $P$.\\
\indent For the numerical integration of the equilibrium equations we used the public RNS code~\cite{rns} by Stergioulas and Friedman~\cite{Stergioulas-1995} (This code is based on the method developed by Komatsu, Eriguchi and Hachisu~\cite{Komatsu-1989} and modifications introduced by Cook, Shapiro and Teukolsky~\cite{Cook-94a}).

\subsection{Keplerian frequency}\label{secsub:3_1}
The derivation of the  Keplerian frequency, in which a rotating star would shed matter at its equator, is a complicated problem. In Newtonian theory has its origin on the balance between gravitational and centrifugal forces and takes a very simple form. However, in General Relativity (GR) exhibits a more complicated dependence on the structure of the star through the interior metric as it is expressed as a self-consistency condition that must be satisfied by the solution to Einstein's  equations.\\
\indent It has been shown by Friedman {\it et al.} ~\cite{Friedman-88} that the turning-point method, which is leading to the points of secular instability, can also be used in the case of uniformly rotating neutron stars. With this consideration, in a constant angular momentum sequence, the turning-point of a sequence of configurations with increasing central density, separates the secular stable from unstable configuration and consequently, the condition
\begin{equation}
\frac{\partial M(\varepsilon_{c},J)}{\partial \varepsilon_{c}}\Bigg\vert_{J={\rm constant}}=0
\label{cond-1}
\end{equation}
where $\varepsilon_{c}$ is the energy density in the center of the neutron star and $J$ is the angular momentum, defines the possible maximum gravitational mass. In general, gravitational (gr) and rest mass (rm) are defined as~\cite{Haensel-07}
\begin{eqnarray}
	M_{\rm gr} &=& \int_{0}^{\rm R} 4\pi r^{2} \epsilon(r) \mathrm{d}r\\
	M_{\rm rm} &=& m_{A} \int_{0}^{\rm R} 4 \pi r^{2} \frac{n(r)}{\left(1-\frac{2GM(r)}{c^{2}r}\right)^{1/2}} \mathrm{d}r
\end{eqnarray}
where $m_{A}$ is the baryonic mass and $n(r)$ is the baryon number density.\\
\indent The absence of analytical solutions for rotating neutron stars leads to numerical estimations for the Keplerian frequency. A significant number of empirical formulas for the Keplerian frequency  had
been produced along the years. The formula is given by~\cite{Haensel-09,Haskell-018}
\begin{equation}
f_{k}=\mathcal{C}_{\alpha} \left(\frac{M_{\rm max}^{\alpha}}{M_{\odot}}\right)^{1/2}\left(\frac{10 km}{R_{\rm max}^{\alpha}}\right)^{3/2} = \mathcal{C}_{\alpha}x_{\rm max}^{\alpha} \text{ } ({\rm Hz})
\label{eq:f_max}
\end{equation}
where
\begin{equation}
x_{\rm max}^{\rm \alpha} = \left(\frac{M_{\rm max}^{\rm \alpha}}{M_{\odot}}\right)^{1/2} \left(\frac{10 km}{R_{\rm max}^{\rm \alpha}}\right)^{3/2}
\label{eq:x}
\end{equation}
and $\alpha$ ($\rm st:static$, $\rm rot:rotating$, $\rm rm;rot: rest$ $\rm mass$ $\rm at$ $\rm rotating$ $\rm configuration$) takes the form of the corresponding configuration. Although this relation is well established, the unknown parameter ($\mathcal{C}_{\alpha}$) depends highly on the various approximations and of course the selected EoSs.\\
\indent It is worth pointing out that while the maximum rotation rate is an increasing function of the EoS's softness, the maximum mass is a decreasing one (considering a fixed mass). Therefore, for a fixed gravitational mass $M$ the softer EoS predicts the lower value of the radius $R$ and consequently, leads to higher values for $f_k$. The latter it had been already noticed by Lattimer {\it et al.}~\cite{Lattimer-1990}. These two constraints restrict the EoS in a narrow region. The above statement is one of the main subjects of the present work.

\subsubsection{The Keplerian frequency, the maximum mass and the corresponding radius of non-rotating neutron stars}\label{secsubsub:3_1_1}

\begin{figure*}
	\centering
	\includegraphics[width=0.49\textwidth]{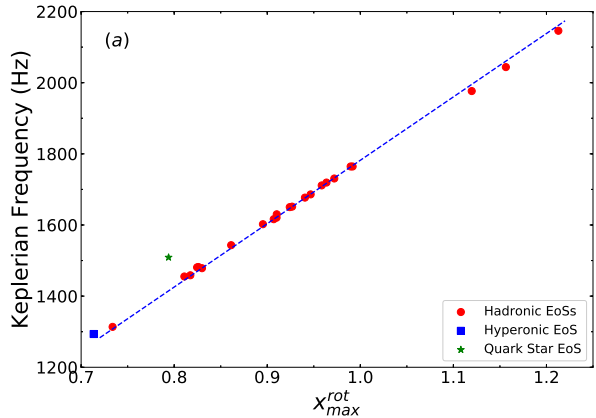}
	~
	\includegraphics[width=0.49\textwidth]{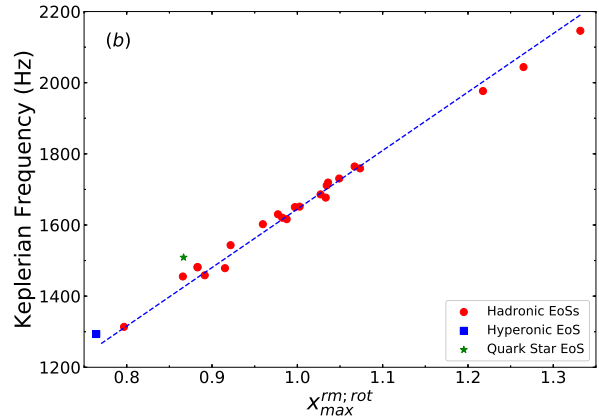}
	\caption{(color online) Keplerian frequency dependence on the quantity (a) $x_{\rm max}^{\rm rot}$ and (b) $x_{\rm max}^{\rm rm;rot}$ for the 23 EoSs (for more details see Eq.~\ref{eq:x}). Blue dashed lines correspond to the best linear trend that fits the data. The data from 23 hadronic EoSs are also presented with red circles. The hyperonic EoS is indicated with the blue square and the quark star EoS with the green star.}
	\label{fig:fgrrm}
\end{figure*}

\begin{figure}[H]
	\includegraphics[width=0.5\textwidth]{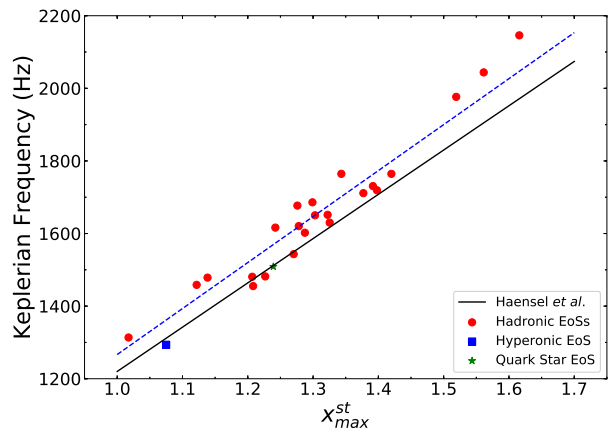}
	\caption{(color online) Keplerian frequency dependence on the quantity $x_{\rm max}^{\rm st}$ for the 23 EoSs (for more details see Eq.~\ref{eq:x}). Blue dashed line corresponds to the best linear trend that fits the data. The data from 23 hadronic EoSs are also presented with red circles. The hyperonic EoS is indicated with the blue square and the quark star EoS with the green star. The black solid line marks the work of Haensel {\it et al.}~\cite{Haensel-09}.}
	\label{fig:fkgm}
\end{figure}

We studied the Keplerian frequency in correlation with the bulk properties of a non-rotating neutron star and specifically on its gravitational mass and the corresponding radius at the maximum mass configuration, using Eq.~\eqref{eq:f_max} and $\rm \alpha = st$.\\
\indent In Fig.~\ref{fig:fkgm} we present the relation~\eqref{eq:f_max} with the corresponding parametrization, which can be found in Table~\ref{tab:2}, updating with this way the work of Haensel {\it et al.}~\cite{Haensel-09}. The value of the parameter $\mathcal{C}_{st}$ is in very good agreement with the current EoSs to a linear term.

\subsubsection{The Keplerian frequency, the maximum mass and the corresponding radius of maximally-rotating neutron stars}
\label{secsubsub:3_1_2}
An interesting relation is also the one between the Keplerian frequency and the macroscopic properties of maximally-rotating neutron stars (maximum gravitational mass and the corresponding radius). Using Eq.~\eqref{eq:f_max} and $\rm \alpha = rot$, it is remarkable that in this scenario, as Fig. \ref{fig:fgrrm}a shows, the linear fit between these quantities ($f_{k}$, $x_{\rm max}^{\rm rot}$) leads to nearly perfect results. The parametrization can be found in Table~\ref{tab:2}.

\subsubsection{The Keplerian frequency, the maximum rest mass and the corresponding radius of maximally-rotating neutron stars}
\label{secsubsub:3_1_3}
In the macroscopic properties of a neutron star, rest mass plays an important role. In order to understand the effects of the rest mass on the Keplerian sequence, we studied the Keplerian frequency dependence on the rest mass and the corresponding radius using Eq.~\eqref{eq:f_max} and $\rm \alpha = rm;rot$.\\
\indent In Fig.~\ref{fig:fgrrm}b we can see the almost linear relation that holds on between these two quantities ($f_{k}$, $x_{\rm max}^{\rm rm;rot}$), enhancing with this way the existence of a relation between rest mass and gravitational mass in neutron stars at the Keplerian frequency. The parametrization can be found in Table~\ref{tab:2}.

\begin{table}[H]
	\squeezetable
	\caption{Parametrization of Eq.~\eqref{eq:f_max} for the different configurations. The relative error (r.e.) between the data and fits is also presented.}
	\begin{ruledtabular}
		\begin{tabular}{cccc}
			$\alpha$ & $\mathcal{C}_{\alpha}$ & r.e.\% & $\mathcal{C}_{\rm Haensel}$\\
			\hline
			st & 1266.68 & 5.6 & 1220 \\
			rot & 1781.90 & $\leq$1 & -- \\
			rm;rot & 1644.75 & 2.2 & -- \\
		\end{tabular}
	\end{ruledtabular}
	\label{tab:2}
\end{table}

\subsubsection{Rest mass and gravitational mass at the maximum mass configuration of maximally-rotating neutron stars}\label{secsubsub:3_1_4}
\begin{figure}[H]
\includegraphics[width=0.5\textwidth]{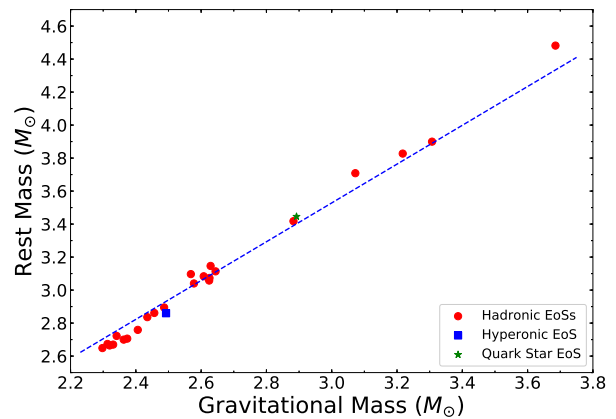}
\caption{(color online) Rest mass dependence on the gravitational mass of a maximally-rotating neutron star at the maximum mass configuration. Blue dashed line corresponds to the best linear trend that fits the data. The data from 23 hadronic EoSs are also presented with red circles. The hyperonic EoS is indicated with the blue square and the quark star EoS with the green star.}
\label{fig:rmgm}
\end{figure}

As a follow-up to Section \ref{secsubsub:3_1_3}, we studied the rest mass dependence on the gravitational mass at the maximum mass configuration for the Keplerian frequency. In Fig. \ref{fig:rmgm}, we can see the almost linear relation between these two quantities, as expected from Section \ref{secsubsub:3_1_3}.

The relation which describes our data is given via the form

\begin{equation}
\left(\frac{M_{\rm max}^{\rm rm;rot}}{M_{\odot}}\right) = 1.17 \left(\frac{M_{\rm max}^{\rm gm;rot}}{M_{\odot}}\right)
\end{equation}

\noindent (the maximum possible error is less than 3.3\%) concluding with this way that the percentage difference between these quantities is around 17\%.

\begin{figure*}
    \centering
		\includegraphics[width=0.49\textwidth]{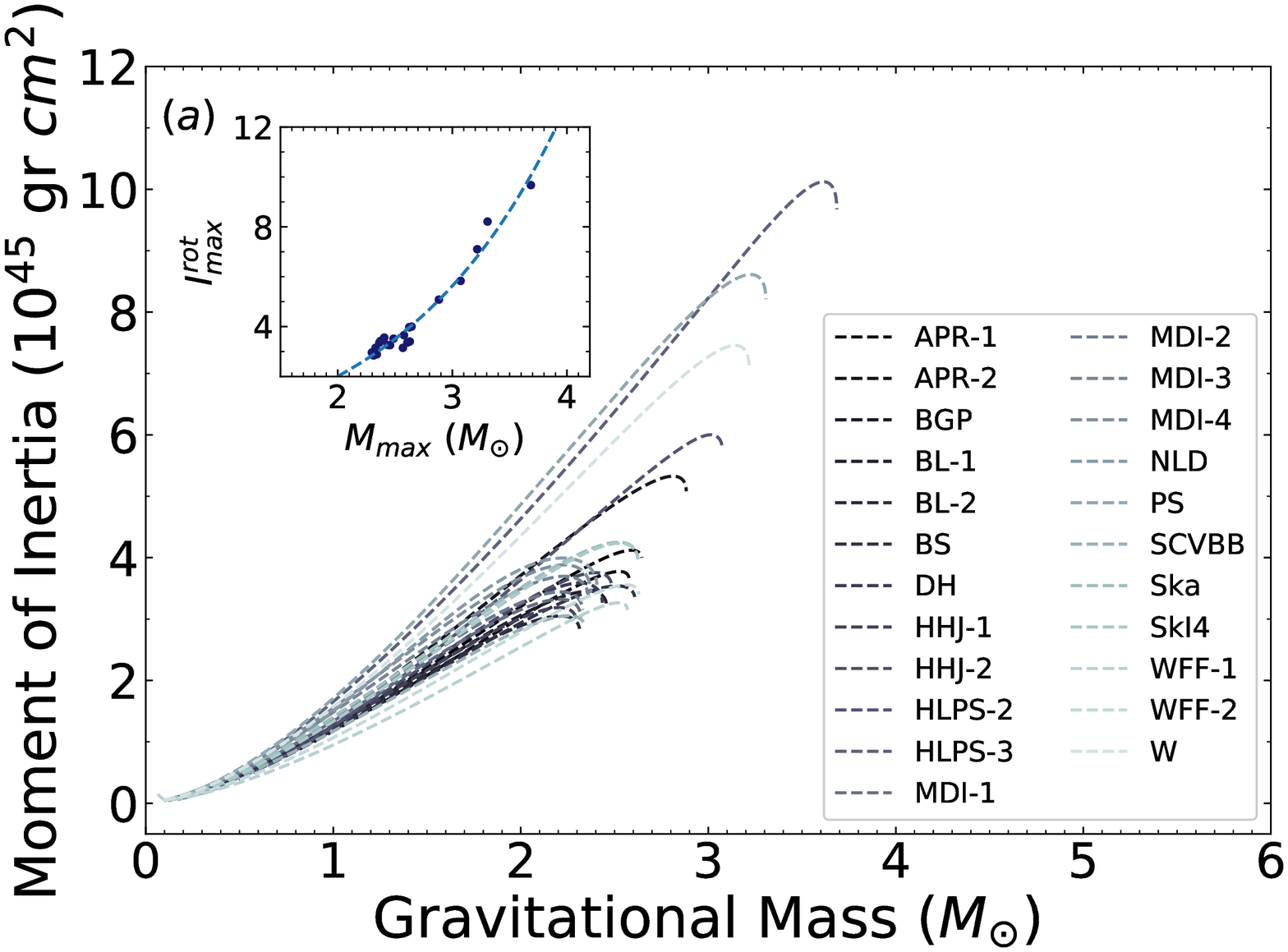}
    ~
        \includegraphics[width=0.49\textwidth]{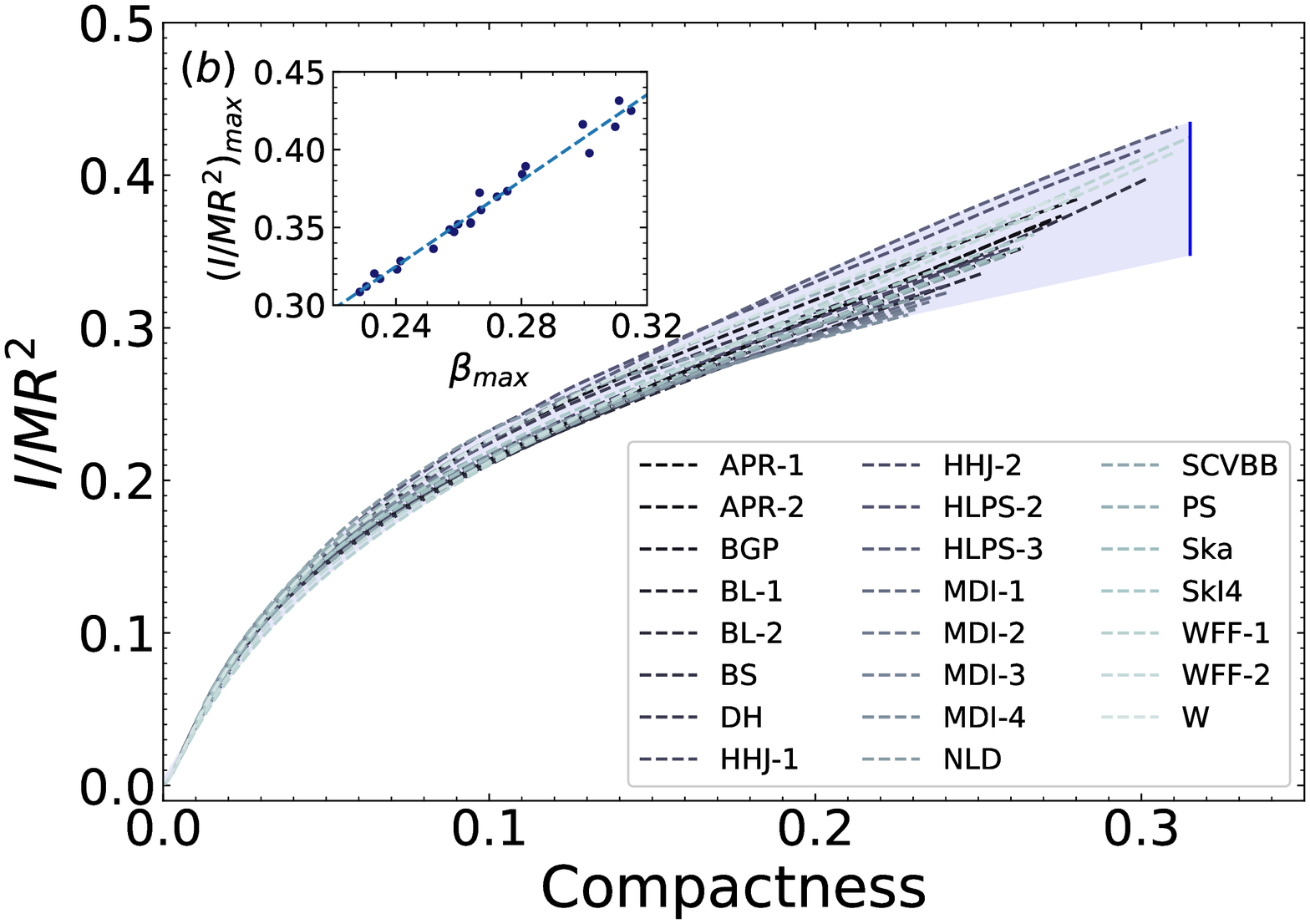}
    \caption{(color online) Moment of inertia dependence (a) on the gravitational mass and (b) on the compactness parameter of a maximally-rotating neutron star for the 23 EoSs. Blue dashed lines in the inside figures correspond to the best fit in each case. The data at the maximum mass configuration are also presented with blue circles in the inside figures.}
    \label{fig:moi}
\end{figure*}

\subsection{Moment of Inertia and Eccentricity}\label{secsub:3_2}
Rotating neutron stars can provide us with more quantities than non-rotating ones that we could study. Among them, there is the moment of inertia and eccentricity. Both these quantities can give us information about the deformation of the mass while its spinning.\\
\indent The moment of inertia~\cite{Stergioulas-03, Cipolletta-015}, which have a prominent role in pulsar analysis, is defined as
\begin{equation}
I=\frac{J}{\Omega}
\end{equation}
where $J$ is the angular momentum and $\Omega$ is the angular velocity. This property of neutron stars quantifies how fast an object can spin with a given angular momentum.\\
\indent We studied the moment of inertia dependence on the gravitational mass for the Keplerian sequence. From Fig.~\ref{fig:moi}a we can see that all EoSs present similar behavior. For this reason, inside Fig. \ref{fig:moi}a, we plotted the moment of inertia values corresponding to maximum mass configuration versus the corresponding gravitational mass. A relation, given by the formula
\begin{equation}
I_{\rm max}^{\rm rot} = -1.568 + 0.883 \exp\left[0.7 \left(\frac{M_{\rm max}^{\rm gm;rot}}{M_{\odot}}\right)\right] \left( 10^{45} \text{ } \rm gr \text{ }cm^{2}\right)
\end{equation}
describes with high accuracy our data, concluding with this way that moment of inertia, at the maximum mass configuration for the Keplerian frequency, can provide us with a universal relation between moment of inertia and the corresponding gravitational mass.

\indent We also studied the dimensionless moment of inertia dependence on the corresponding compactness parameter~\cite{Lattimer-05a}, which, in general, it is defined as

\begin{equation}
\beta = \frac{GM}{Rc^{2}}
\end{equation}
where $R$ corresponds to the equatorial radius of neutron star.

In Fig.~\ref{fig:moi}b we present a window where moment of inertia and compactness parameter can lie (shadowed/gray region), constraining with this way both these quantities. There is an empirical relation, derived from the data, that can describe this window. The form of this empirical relation is
\begin{equation}
I/ M R^{2} = \alpha_{1} + \alpha_{2} \beta + \alpha_{3} \beta^{2} + \alpha_{4} \beta^{3} + \alpha_{5} \beta^{4}
\label{eq:imr2}
\end{equation}
where the coefficients for the two edges are shown in Table \ref{tab:3}. It is clear from Fig. \ref{fig:moi}b and Eq. \eqref{eq:imr2} that if we have a measurement of moment of inertia, or compactness parameter, we could extract the interval where the other parameter can lie.\\
\indent As a consequence, by constraining simultaneously these two quantities, we could impose constraints on the radius of neutron stars, which still remains an open problem.

\begin{table}[H]
\squeezetable
\caption{Coefficients of the empirical relation~\eqref{eq:imr2} for the two edges of the window presented in Fig. \ref{fig:moi}b.}
\begin{ruledtabular}
\begin{tabular}{cccccc}
Edges & $\alpha_{1}$ & $\alpha_{2}$ & $\alpha_{3}$ & $\alpha_{4}$ & $\alpha_{5}$ \\
\hline
Upper & 0.005 & 4.01 & -24.79 & 86.66 & -110.33 \\
Lower & 0.005 & 3.38 & -17.45 & 49.68 & -55.36 \\
\end{tabular}
\end{ruledtabular}
\label{tab:3}
\end{table}

\begin{figure*}
	\centering
	\includegraphics[width=0.4\textwidth]{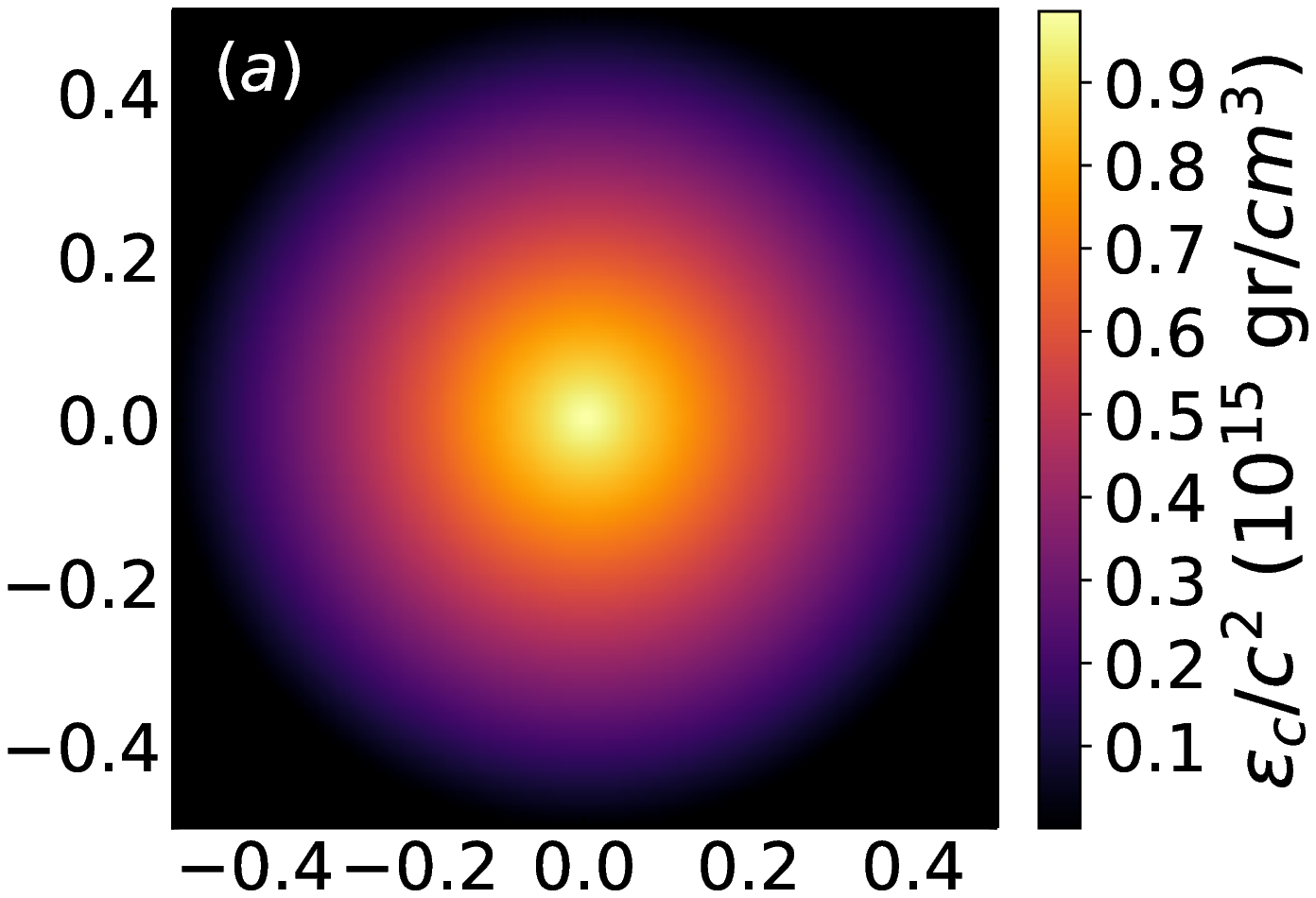}
	~
	\includegraphics[width=0.4\textwidth]{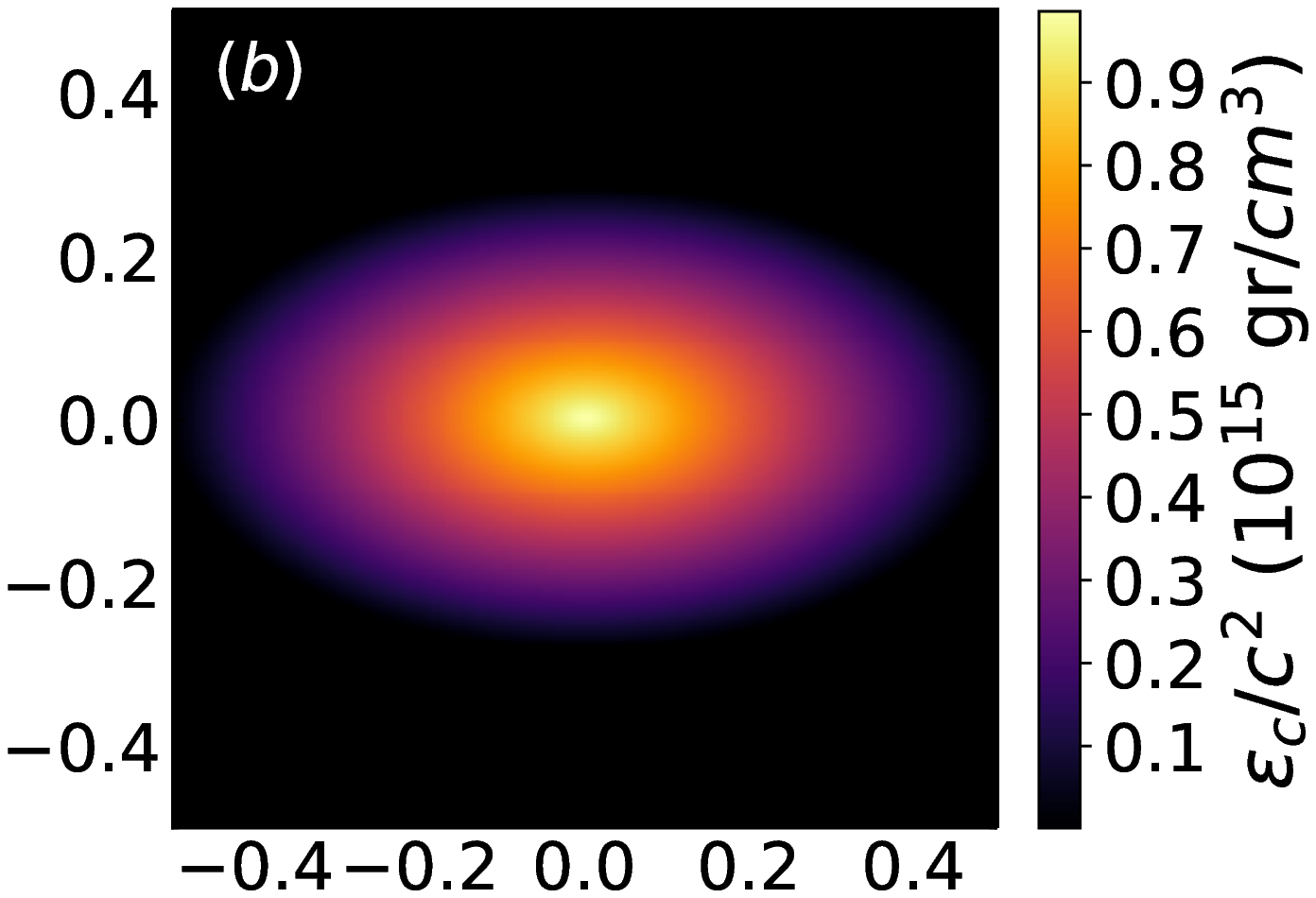}
	\caption{(color online) Contours of constant density of a neutron star model with central density equals to $10^{15}$ $\rm gr $ $\rm cm^{-3}$, both (a) in the non-rotating case and (b) in the rotating one with frequency equals to the Keplerian frequency for the APR-1 EoS. The axis had been scaled in a way that the maximum radius corresponds to 0.5.}
	\label{fig:ed}
\end{figure*}

\begin{figure}[H]
	\includegraphics[width=0.5\textwidth]{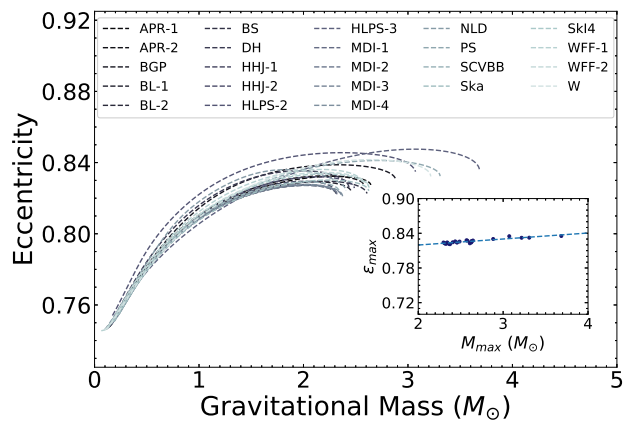}
	\caption{(color online) Eccentricity dependence on the gravitational mass of a maximally-rotating neutron star. (Inside) The eccentricity values as a function of the corresponding gravitational mass at the maximum mass configuration. Blue dashed line in the inside figure corresponds to the best fit. The data at the maximum mass configuration are also presented with blue circles in the inside figure.}
	\label{fig:ecc}
\end{figure}

From Fig. \ref{fig:moi}b we can see that all EoSs present similar behavior. For this reason, inside Fig. \ref{fig:moi}b, we plotted the dimensionless moment of inertia values corresponding to maximum mass configuration versus the corresponding compactness parameter. A relation, given by the formula
\begin{equation}
	\left(I/MR^{2}\right)_{\rm max}  = -0.006 + 1.379 \beta_{\rm max}
\end{equation}
describes with high accuracy our data, concluding with this way that dimensionless moment of inertia, at the maximum mass configuration for the Keplerian frequency, can provide us with a universal relation between dimensionless moment of inertia and the corresponding compactness parameter.\\
\indent Eccentricity, is the main quantity that is related to the deformation of the star. Rapid rotation deforms the models of equilibrium and in order to see how these models change we calculate the eccentricity, which is given by the form~\cite{Cipolletta-015}
\begin{equation}
\epsilon = \sqrt{1-\left(\frac{r_{\rm pol}}{r_{\rm eq}}\right)^{2}}
\end{equation}

\noindent where the $r_{\rm pol}$ and $r_{\rm eq}$ are the polar and equatorial radius of the star, respectively.\\
\indent For a schematic presentation of the energy inside a neutron star, in Fig.~\ref{fig:ed} we present the contours of constant density of a neutron star model with central density equals to $10^{15}$ $\rm gr \text{ } cm^{-3}$, both in the non-rotating case and in the rotating one with frequency equals to the Keplerian frequency. For the sake of example we used the APR-1 EoS.\\
\indent Performing the same analysis as for moment of inertia, we studied the eccentricity dependence on the gravitational mass for the Keplerian sequence and the eccentricity values corresponding to maximum mass configuration on the corresponding gravitational mass, as Fig.~\ref{fig:ecc} shows. A relation, given by the formula
\begin{equation}
\epsilon_{\rm max} = 0.799 + 0.01 \left(\frac{M_{\rm max}}{M_{\odot}}\right)
\label{eq:eccmax}
\end{equation}
describes with high accuracy our data, concluding with this way that eccentricity, at the maximum mass configuration for the Keplerian frequency, is an EoS-independent property.

\subsection{The Kerr parameter}\label{secsub:3_3}
The Kerr space-time provided from the Einstein's field equations, give us the so-called Kerr black holes \cite{Cipolletta-015, Lo-2011}. These rotating black holes can be fully described from the gravitational mass (M) and the angular momentum (J). In order to have a meaningful Kerr black hole, the relation $J\ge GM^{2}/c$ (Kerr bound) must hold, or otherwise, we have a naked singularity. A naked singularity is a black hole without a horizon and can be considered as closed timelike curves, where causality would be violated. While there is no rigorous proof from Einstein's field equations, the cosmic-censorship conjecture implies that a generic gravitational collapse cannot form a naked singularity~\cite{Virbhadra-2000,Virbhadra-2002,Virbhadra-2008,Virbhadra-2009}. This is the reason why the astrophysical black holes should satisfy the Kerr bound~\cite{Cipolletta-015, Lo-2011}.\\
\indent The gravitational collapse of a massive rotating neutron star, constrained to mass-energy and angular momentum conservation, creates a black hole with almost the same mass and angular momentum as the prior neutron star. In this case, an important quantity to study, directly related with black holes as well as neutron stars, is the dimensionless angular momentum~\cite{Chakrabarti-2014}, which is defined as
\begin{equation}
j=\frac{cJ}{GM_{\odot}^{2}}
\end{equation}

\noindent and it is known as dimensionless spin parameter. As a consequence of this parameter, we can define a new one, starting from the parameter $\alpha$, which is the angular momentum in units of mass and it is given by the form~\cite{Paschalidis-2017}
\begin{equation}
\alpha \equiv \frac{J}{M} = j \frac{GM_{\odot}^{2}}{c}\frac{1}{M}
\label{eq:k_a}
\end{equation}

As a follow, using Eq.~\eqref{eq:k_a}, the well-known \textit{Kerr parameter} takes the form
\begin{equation}
\mathcal{K} = \frac{\alpha}{M} \frac{c}{G} = j \left(\frac{M_{\odot}}{M}\right)^{2}
\end{equation}

\begin{figure}[H]
	\includegraphics[width=0.5\textwidth]{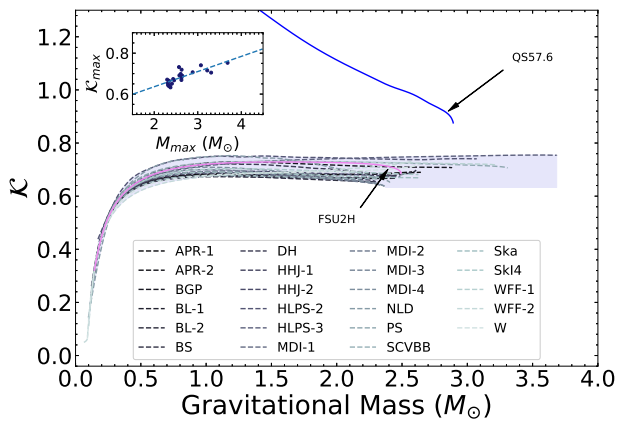}
	\caption{(color online) Kerr parameter dependence on the gravitational mass of a maximally-rotating neutron star. (Inside) The Kerr parameter values as a function of the corresponding gravitational mass at the maximum mass configuration. The blue dashed line in the inside figure corresponds to the best linear trend. The data at maximum mass configuration for the 23 hadronic EoSs are presented with blue circles in the inside figure. FSU2H and QS57.6 EoSs are also indicated with the two solid lines.}
	\label{fig:kerr}
\end{figure}

\noindent  The dependence of this parameter on the gravitational mass at the Keplerian sequence can be seen in Fig.~\ref{fig:kerr}.

\indent Although the meaning of this parameter at black-holes physics is so interesting and fundamental (there is a maximum value at 0.998~\cite{Thorne-1974}), that's also the case for other compact objects such as neutron stars. In order to find a way to constrain the value of the Kerr parameter in neutron stars, we studied the dependence of this parameter on the total gravitational mass for the Keplerian sequence. From Fig. \ref{fig:kerr}, we can see that the maximum value of the Kerr parameter for neutron stars is around 0.75. While there is a number of EoSs that hold on near this value, the maximum value achieved from HLPS-3. This EoS is the stiffest equation that we have and produces maximum mass greater than all the others. Strictly speaking, if we consider this EoS as the one that produces the maximum possible mass in the maximum mass configuration at the Keplerian sequence, then we could constrain the maximum value of the Kerr parameter in neutron stars.\\
\indent In Fig.~\ref{fig:kerr} we present also a window (shadowed/gray region) where the Kerr parameter can lie. There is an empirical relation, derived from the data, that can describe this window. The form of this empirical relation is
\begin{equation}
\mathcal{K} = d_{1} + d_{2}\coth\left[d_{3}\left(\frac{M_{\rm max}}{M_{\odot}}\right)\right]
\label{eq:w_kerr}
\end{equation}
\noindent where the coefficients for the two edges are shown in Table \ref{tab:4}. It is clear from Fig. \ref{fig:kerr} and Eq.~\eqref{eq:w_kerr} that if we have a measurement of gravitational mass, or spin parameter, we could extract the interval where the other parameter can lie.\\
\indent As a consequence, by measuring accurately and simultaneously these two quantities, we could impose constraints on the EoS.

\begin{table}[H]
\squeezetable
\caption{Coefficients of the empirical relation~\eqref{eq:w_kerr} for the two edges of the window presented in Fig. \ref{fig:kerr}.}
\begin{ruledtabular}
\begin{tabular}{cccc}
Edges & $d_{1}$ & $d_{2}$ & $d_{3}$ \\
\hline
Upper & 0.86 & -0.12 & 1.54 \\
Lower & 0.86 & -0.21 & 2.67 \\
\end{tabular}
\end{ruledtabular}
\label{tab:4}
\end{table}

\indent In addition, in Fig. \ref{fig:kerr}, we plotted the maximum values of the Kerr parameter versus the corresponding gravitational mass. It seems that a linear relation holds between these two quantities, given by the equation
\begin{equation}
\mathcal{K}_{\rm max} = 0.488 + 0.074 \left(\frac{M_{\rm max}}{M_{\odot}}\right)
\end{equation}
There are two important reasons for constraining the Kerr parameter at neutron stars: First, the existence of a maximum value at the Kerr parameter, can lead to possible limits for the compactness on neutron stars; strictly speaking, the maximum value of the Kerr parameter for neutron stars implies a maximum value on the possible maximum mass of rotating neutron stars in the universe and second, can be a criteria for determining the final fate of the collapse of a rotating compact star~\cite{Lo-2011}.

\begin{figure*}
	\centering
	\includegraphics[width=0.49\textwidth]{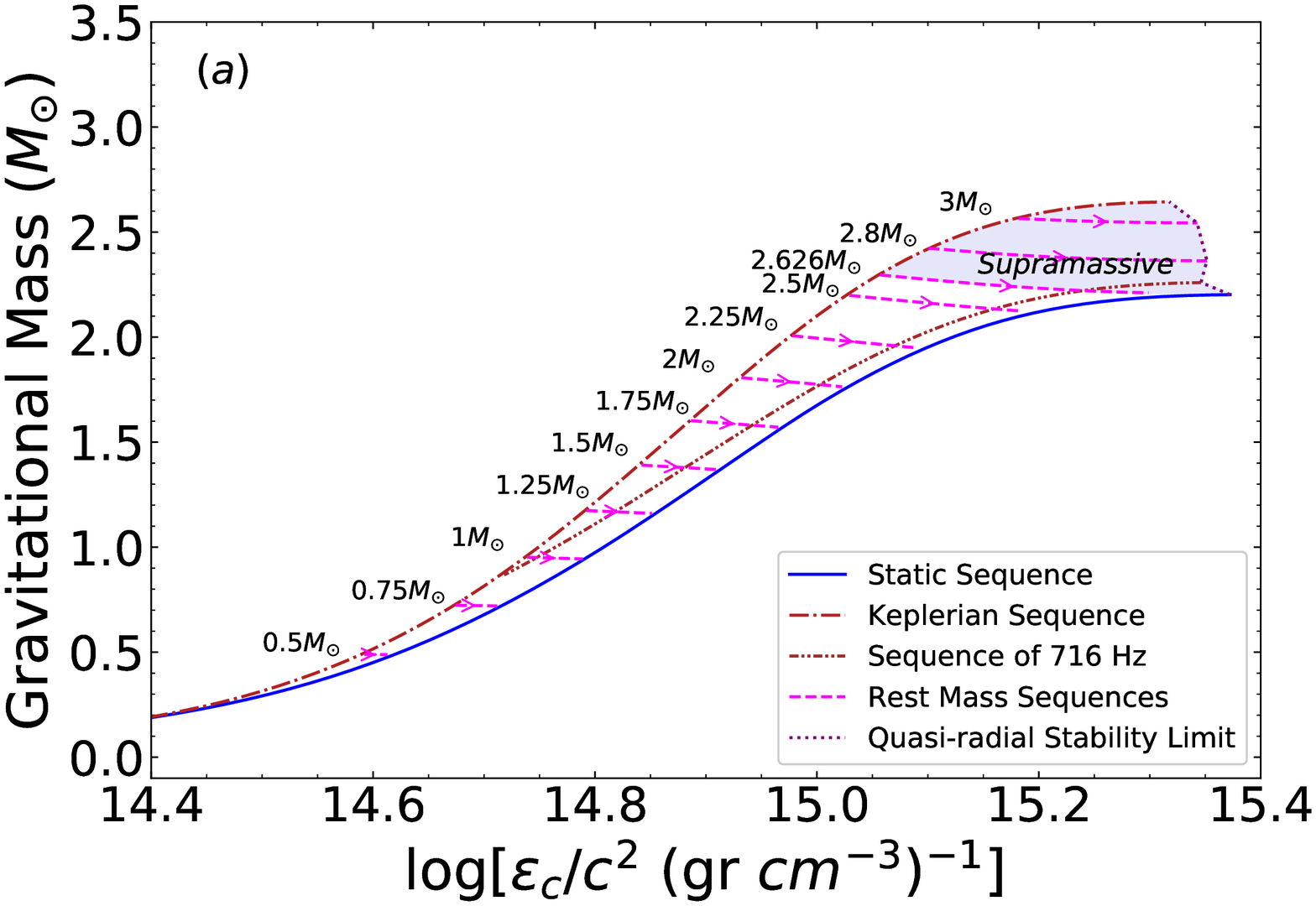}
	~
	\includegraphics[width=0.49\textwidth]{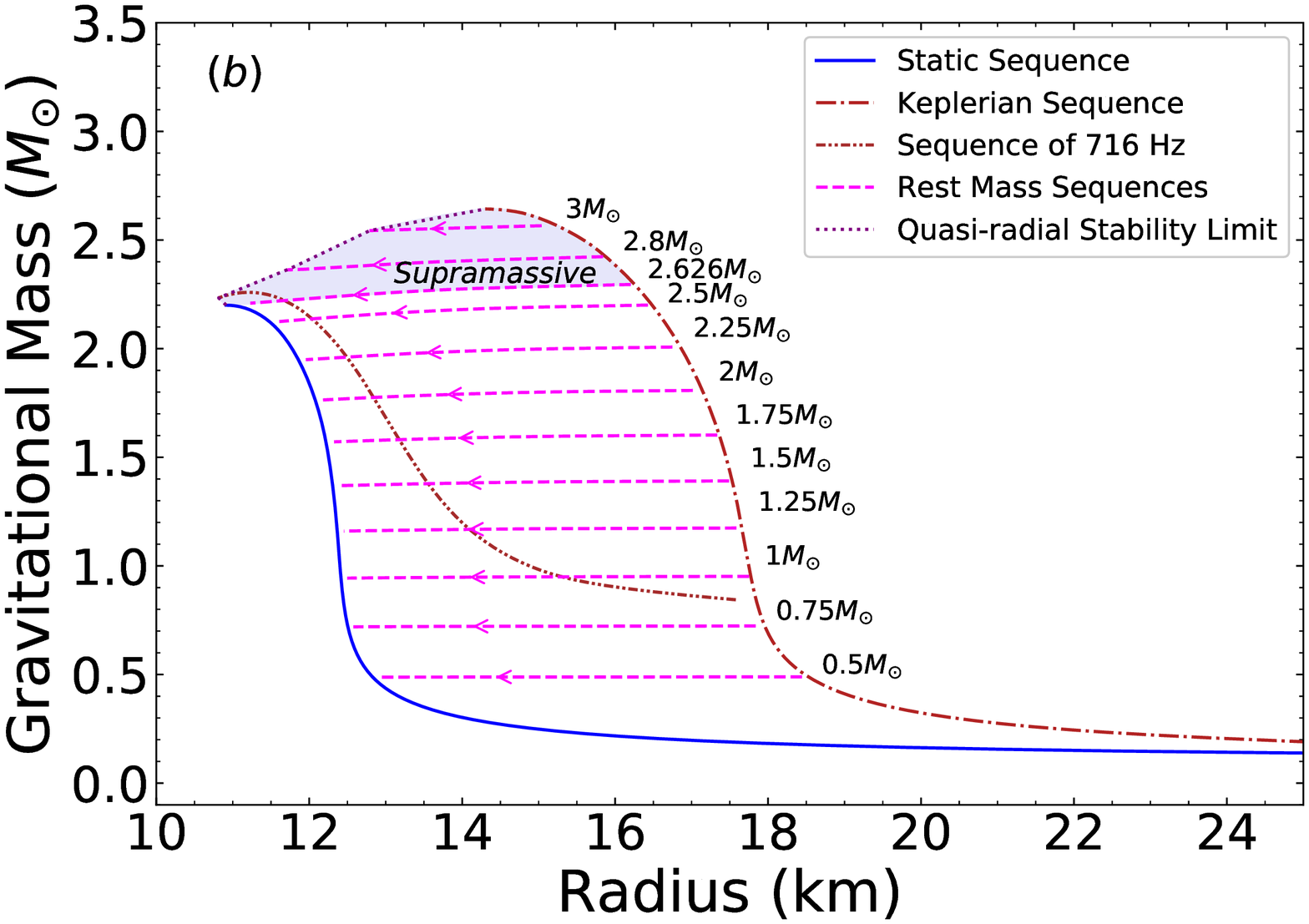}
	\caption{(color online) Normal and supramassive evolutionary sequences of constant rest mass are presented as the dependence of the (a) gravitational mass on the central energy density and (b) gravitational mass on the corresponding radius for the APR-1 EoS. Non-rotating case is presented with the blue solid curve while the maximally-rotating one with the red dashed-dotted curve. Constant rest mass sequences are presented with the fuchsia dashed lines, where the rest mass value is also noted. The 716 Hz limit is also presented with the brown dashed-dotted-dotted curve. The quasi-radial stability limit is presented with the purple dotted line.}
	\label{fig:rm}
\end{figure*}

\indent Finally, it is clear from Fig.~\ref{fig:kerr} that the Kerr parameter dependence on the gravitational mass of a quark star is quite different than the one on neutron stars. The Kerr parameter of quark stars can be significantly larger than the maximum value of this parameter on neutron stars. In case of the hypeonic EoS, the dependence between gravitational mass and the Kerr parameter exhibits similar behavior with the hadronic ones. A detail study in this direction is in progress.

\subsection{Constant rest mass sequences}\label{secsub:3_4}
The rest mass sequences, also called as time evolutionary sequences, based on an EoS, are roughly horizontal lines that extend from the Keplerian sequence to the non-rotating end point or at the axisymmetric instability limit~\cite{Cook-94a,Cook-94b,Cook-94c}. The latter depends only on the rest mass value of the selected EoS. For a given EoS, the sequences that are below the rest mass value that corresponds to the maximum mass configuration at the non-rotating model, they have a non-rotating member, and as a consequence, are stable and terminate at the non-rotating model sequence. Above this value, none of the sequences have a non-rotating member. Instead, they are unstable and terminate at the axisymmetric instability limit. The onset that extends from the maximum mass point on the non-rotating limit sequence to the one on the mass-shedding limit sequence is the quasi-radial stability limit. The total region that models are unstable is defined via the rest mass sequence that corresponds to the maximum mass configuration of the non-rotating model, as Fig. \ref{fig:rm} shows (shadowed/gray region). Above this value, the models have masses larger than the maximum mass of the non-rotating model and in that case, are called supramassive models~\cite{Friedman-13}. It should be noted that models to the axisymmetric area of the shadowed/gray region, which is not shown at the corresponding figures, are also unstable.\\
\indent To be more specific, if a neutron star spin-up by accretion and becomes supramassive, then it would subsequently spin-down along the constant rest mass sequence until it reaches the axisymmetric instability limit and collapse to a black hole. There is a case where some relativistic stars could be born as supramassive ones, or even more, become one as a result of a binary merger. In this case, the star would be initially differentially rotating and collapse would be triggered by a combination between spin-down effect and viscosity (the force that driving the star to uniform rotation)~\cite{Friedman-13}.\\
\indent Although the sequence with rest mass corresponding to the maximum mass configuration of the non-rotating model extends to the right area of the quasi-radial stability limit, the unstable one, it is the last one that has a stable part (half of the sequence terminates at the maximum mass configuration of the non-rotating model). While, below this sequence, all the remaining ones are unconditionally stable against gravitational collapse, above this sequence, all sequences would evolve toward catastrophic collapse to a black hole. In Fig. \ref{fig:rm}, we can see that if we have a neutron star with rest mass in the white region, it would evolve towards stable configuration at the non-rotating sequence, but if we have a star in the shadowed/gray region, it would subsequently spin-up and evolve towards catastrophic collapse to a black hole~\cite{Shibata-2003,Lasky-2014,Ravi-2014}. The direction of evolution for constant rest mass sequences is noted with the existence of an arrow on them.

\begin{figure*}
	\centering
	\includegraphics[width=0.49\textwidth]{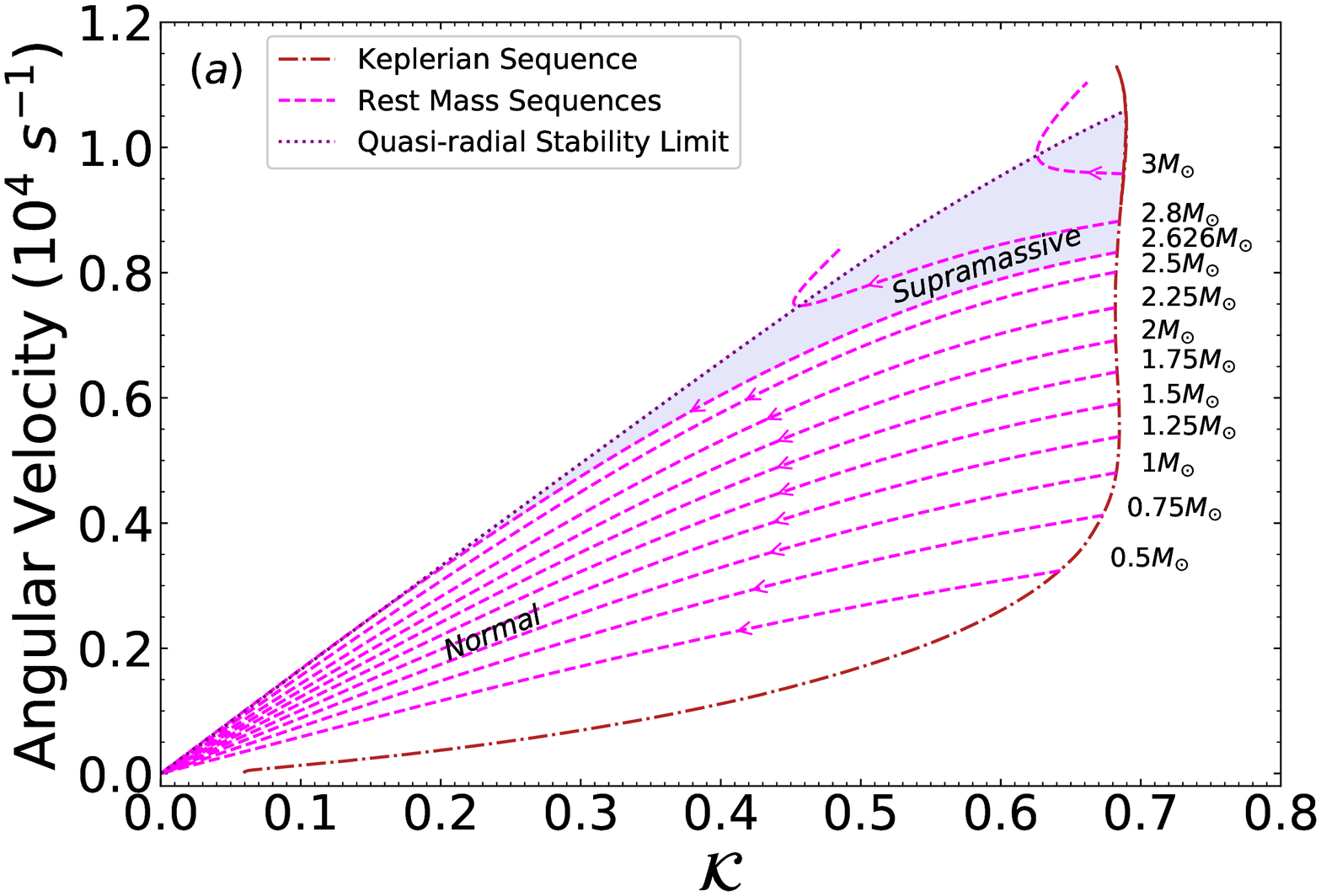}
	~
	\includegraphics[width=0.49\textwidth]{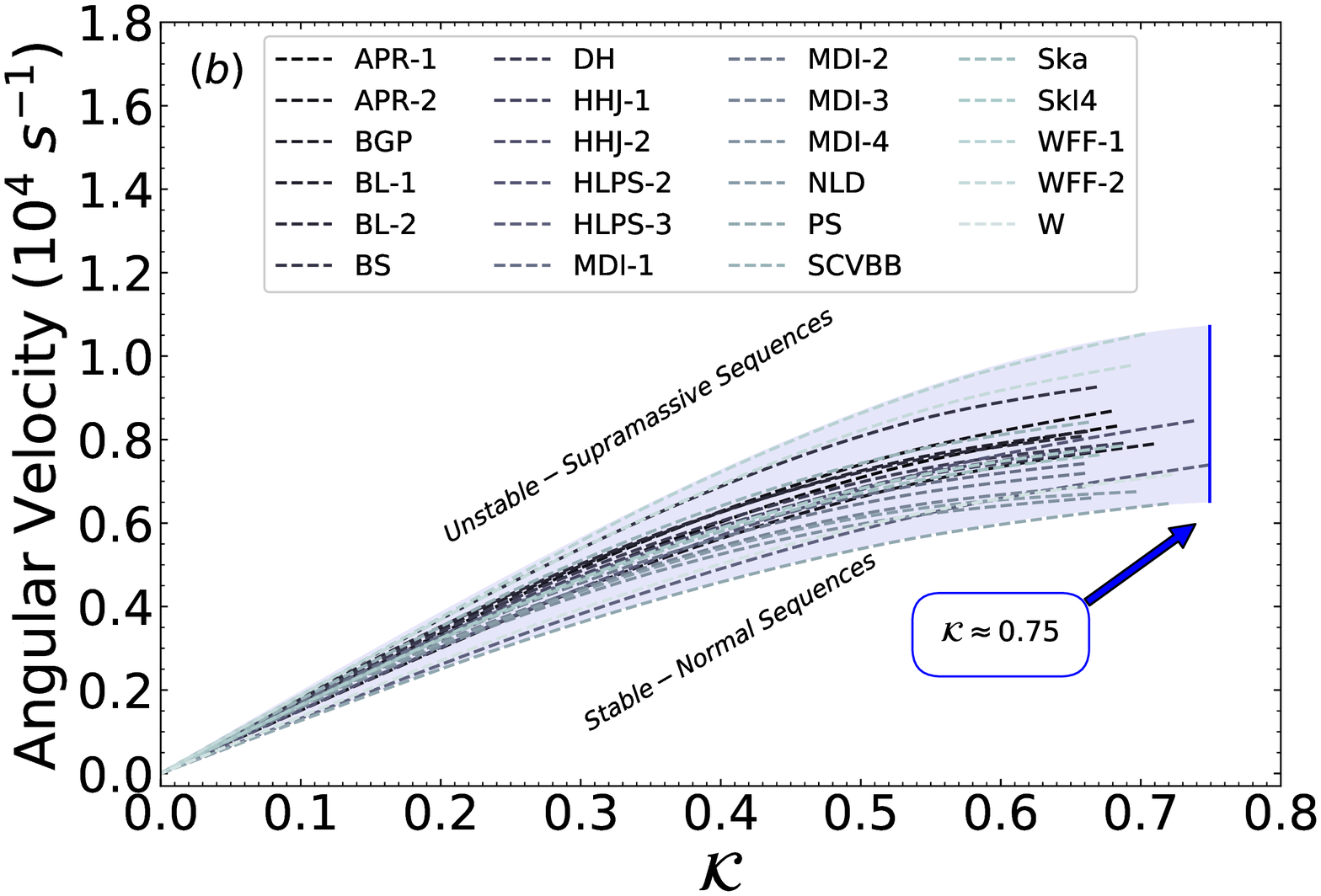}
	\caption{(color online) (a) Normal and supramassive evolutionary sequences of constant rest mass are presented as the dependence of the angular velocity on the Kerr parameter for the APR-1 EoS. Maximally-rotating case is presented with the red dashed-dotted curve while the constant rest mass sequences are presented with the fuchsia dashed curves, where the rest mass values are also noted. The quasi-radial stability limit is presented with the purple dotted curve. (b) Last stable rest mass sequences for the 23 EoSs as the dependence of angular velocity on the Kerr parameter. Supramassive and normal area  are shown to guide the eye. The maximum value of the Kerr parameter is also noted.}
	\label{fig:rm1}
\end{figure*}
In all cases, neutron stars which evolve along normal evolutionary sequences, never spin-up as they lose angular momentum. In contradiction to neutron stars on normal evolutionary sequences, neutron stars on supramassive ones, because their unstable portion is always at higher angular velocity than the stable portion, at the same value of angular momentum, must spin-up with angular momentum loss in the neighborhood of the stability limit. If the neutron star is massive enough, then the evolutionary sequence (supramassive) exhibit an extended region where spin-up is allowed. This effect may provide us an observable precursor to gravitational collapse to a black hole~\cite{Lo-2011,Cipolletta-015}. The latter is shown clearly in Fig. \ref{fig:rm1}a.\\
\indent Following the concept from Fig. \ref{fig:rm1}a, we have constructed the last stable rest mass sequence (LSRMS) for the 23 hadronic EoSs, as shown in Fig. \ref{fig:rm1}b. This sequence is the one that divides the stable from unstable region, or in other words, the normal from supramassive evolutionary sequences. In Fig. \ref{fig:rm1}b, we present a window (shadowed/gray region) where the last stable rest mass sequence can lie and in fact, because the last stable rest mass sequence is the one that corresponds to the maximum mass configuration at the non-rotating model, this is also the region where the EoS can lie, constraining with this way, simultaneously, the spin parameter and the angular velocity.
There is an empirical relation, derived from the data, that can describe this window. The form of this empirical relation is

\begin{equation}
\Omega = \left(b_{1} \mathcal{K} + b_{2} \mathcal{K}^{2} + b_{3} \mathcal{K}^{3}\right) 10^{4} \quad ({\rm s^{-1}})
\label{eq:lsrm}
\end{equation}

\noindent where the coefficients for the two edges are shown in Table \ref{tab:5}. It is clear from Fig. \ref{fig:rm1}b and Eq. \eqref{eq:lsrm} that if we have a measurement of angular velocity, or spin parameter, we could extract the interval where the other parameter can lie.\\
\indent As a consequence, by constraining simultaneously these two quantities, we could significantly narrow the existing area of EoS.

\begin{table}[H]
\squeezetable
\caption{Coefficients of the empirical relation~\eqref{eq:lsrm} for the two edges of the window presented in Fig. \ref{fig:rm1}b.}
\begin{ruledtabular}
\begin{tabular}{cccc}
Edges & $b_{1}$ & $b_{2}$ & $b_{3}$ \\
\hline
Upper & 1.94 & 0.117 & -1.058 \\
Lower & 1.35 & -0.305 & -0.449 \\
\end{tabular}
\end{ruledtabular}
\label{tab:5}
\end{table}

\subsection{Upper bound for density of cold baryonic matter}\label{secsub:3_5}
Although we employ realistic EoSs to solve numerically equilibrium equations in neutron stars, analytical solutions are far from being insignificant. Useful information can be gained by the comparison between solutions of the Einstein's field equations with numerical solutions for different models of EoSs and the analytical solutions \cite{Lattimer-05}. Two classes derive from analytical solutions: (a) normal neutron stars and (b) self-bound neutron stars. In the first case, the energy density vanishes at the surface where the pressure vanishes and in the second one, the energy density is finite at the surface.\\
\indent In this work only the first case scenario will be studied. It is most natural to solve numerically the Tolman–Oppenheimer–Volkoff (TOV, Einstein's equations for a non-rotating spherical symmetric object) equations~\cite{Shapiro-83,Glendenning-2000,Haensel-07}, by introducing an EoS describing the relation between pressure and density which is expected to describe the fluid interior. The other possibility is trying find out analytical solutions of TOV equations with the risk of obtaining solutions without physical interest. Actually, there are hundreds of analytical solutions of TOV equations~\cite{Kramer-1980,Delgaty-1998}. However, just few of them are of physical interest. Moreover, there are only three that satisfy the criteria that the pressure and energy density vanish on the surface of the star and also that they both decrease monotonically with increasing radius. These three solutions are the Tolman VII, the Buchdahl and the Nariai IV~\cite{Moustakidis-2017}. The main difference between these analytical solutions is related with the maximum value of compactness at which they took effect. For example, the  Buchdahl  solution is applicable only for neutron stars with compactness up to the value $\beta=0.17$ and in general produces soft EoSs. The Tolman VII solution leads to even stiffer EoSs and consequently is suitable to describe compact objects with compactness value up to $\beta=0.34$ (for more details see Ref.~\cite{Moustakidis-2017}). The Nariai IV solution exhibits similar behavior with the Tolman VII. In particular the Tolman VII is of great interest since it has the specific property that the pressure and density vanish at the surface of the star. It has been extensively employed to neutron star studies and the details of this analytical solution had been given in Appendix~\ref{app:3}.\\
\indent It has been shown by Lattimer {\it et al.} \cite{Lattimer-05} that the Tolman VII solution forms an absolute upper limit, which confirmed empirically by using a large number of EoSs, in density inside any compact star (see also Ref.~\cite{Lattimer-2011,Zhang-2019}). This is also the case for rotating stars with rotation rates up to the Keplerian (mass-shedding) rate.

\begin{figure}[H]
\includegraphics[width=0.5\textwidth]{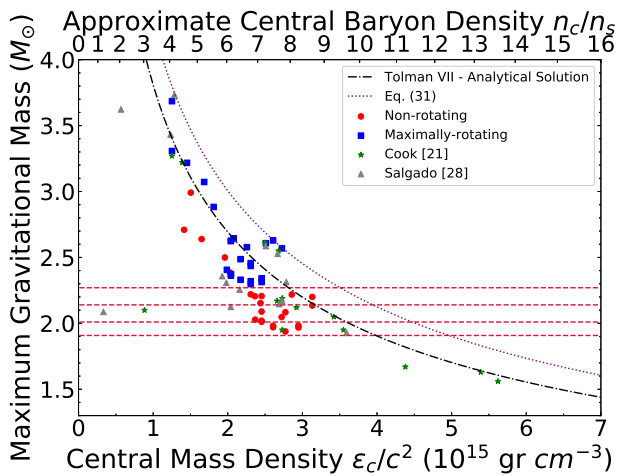}
\caption{(color online) Gravitational mass dependence on the central energy density and the central baryon density at the maximum mass configuration for the 23 EoSs at the non-rotating and maximally-rotating case. Red circles correspond to the non-rotating case, blue squares to the maximally-rotating one, green stars to Cook's~\cite{Cook-94c} data and grey triangles to Salgado's~\cite{Salgado-94b} data. The red horizontal dashed lines correspond to the observed neutron star mass limits ($1.908 M_{\odot}$, $2.01 M_{\odot}$, $2.14 M_{\odot}$ and $2.27 M_{\odot}$). For comparison, the Tolman-VII analytical solution with the black dashed-dotted curve and the Eq.~\eqref{eq:ued} with the purple dotted one are shown.}
\label{fig:mec}
\end{figure}

\indent At that time, the maximum masses of the existed EoSs were fully included in the region under the Tolman VII solution; the same holds for the rotating models. In recent years, new EoSs have been introduced and old ones, that could not describe the maximum observed neutron star mass \cite{Demorest-010,Fonseca-016,Arzoumanian-2018,Antoniadis-013,Cromartie-19,Linares-18}, have been rejected. In this work, using a total of 23 hadronic EoSs that predict the observed maximum neutron star mass \cite{Demorest-010,Fonseca-016,Arzoumanian-2018,Antoniadis-013,Cromartie-19,Linares-18}, we have confirmed that the Tolman VII curve marks the upper limit to the energy density inside a star but without taking into account the rotation (Tolman VII can describe the majority of them). If we add rotation to our models, then this curve is not able to describe anymore the new data as they shift, concerning the plotted area, up and left. For this reason, we propose here, that if there is a curve, like the Tolman VII solution, shifted to the right, that would be a suitable solution to fully describe the maximum energy density inside a star. In other words, the existence of this curve can help to form an absolute upper limit in density inside any compact object.\\
\indent The proposed expression, described by the form
\begin{equation}
\frac{M}{M_{\odot}} = 4.25 \sqrt{\frac{10^{15}\text{ } \rm gr \text{ } cm^{-3}}{\varepsilon_{c}/c^{2}}}
\label{eq:ued}
\end{equation}
\noindent can fully describe both the non-rotating and maximally-rotating configuration. The advantages of having this relation are that a) it can describe the non-rotating configuration having as a guide the corresponding maximally-rotating one (the Tolman VII analytical solution cannot describe all of them, as displayed in Fig.~\ref{fig:mec}), and b) it can also describe the maximally-rotating configuration.\\
\indent In Fig. \ref{fig:mec} we present the results of the 23 hadronic EoSs, for the non-rotating and maximally-rotating case, Cook's~\cite{Cook-94c} and Salgado's~\cite{Salgado-94b} data, Tolman VII analytical solution and the proposed solution \eqref{eq:ued}. The observed neutron star mass limits are also presented to guide the eye.\\
\indent The knowledge of the central density at the maximally-rotating case is important for studying the pulsar's time evolution. In particular, following the spin-down trail of a millisecond pulsar, the central density increases and the highly compressible quark matter will replace the existed nuclear matter. This effect is directly connected to the reduction of moment of inertia. Henceforth, the central density can inform us on the appearance of a phase transition in its interior. The latter can leads to the important back-bending phenomenon in pulsars~\cite{Glendenning-1998}.\\
\indent Another interesting effect that presented via the Fig. \ref{fig:mec}, is the connection that establishes between gravitational mass at the maximum mass configuration and the corresponding central energy density. Besides the fact that can provide us with the absolute upper limit in density inside any compact star, it can also directly connect the macroscopic properties of neutron star with the microscopic ones.

\begin{figure*}
	\centering
	\includegraphics[width=0.32\textwidth]{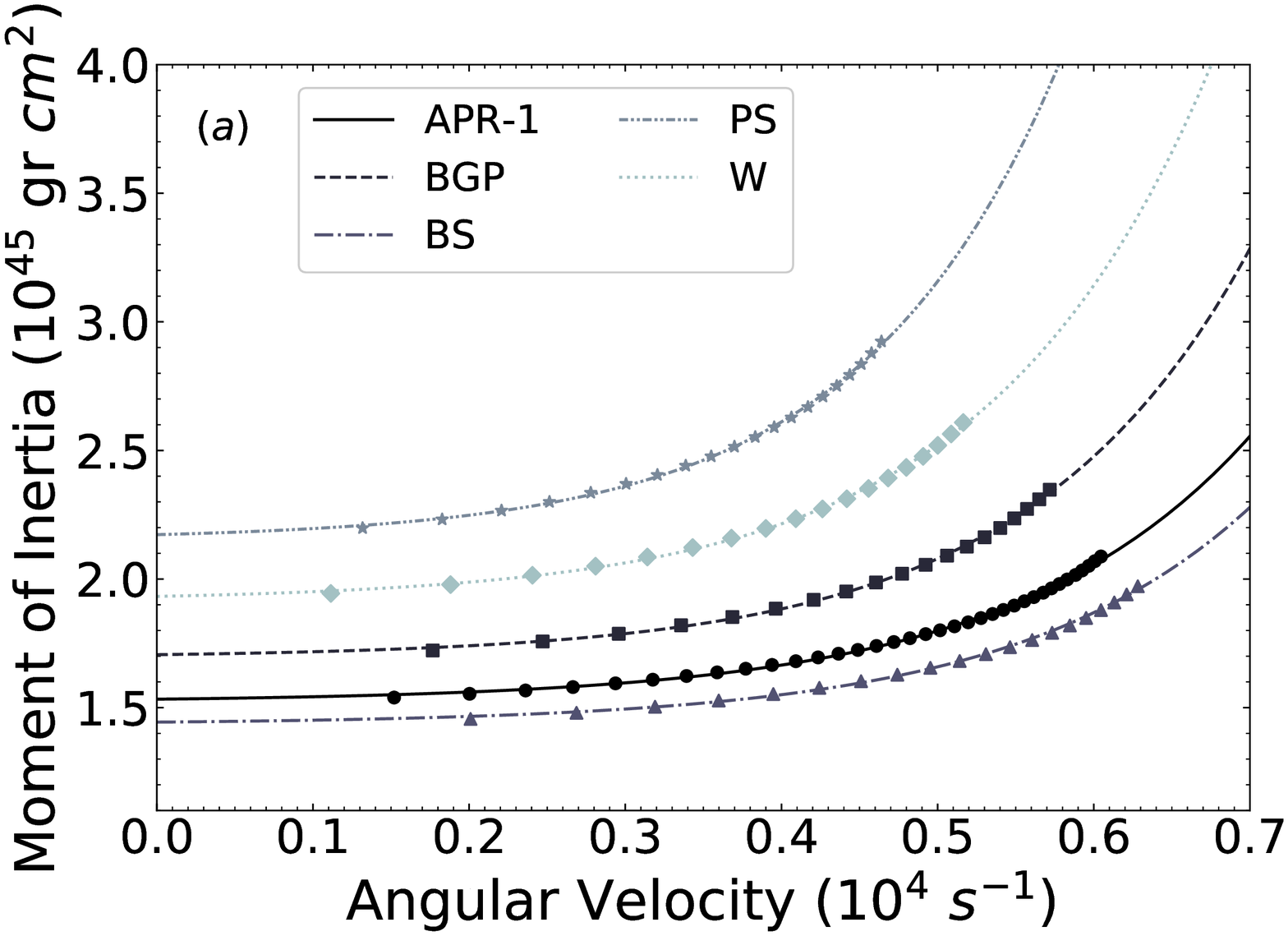}
	~
	\includegraphics[width=0.32\textwidth]{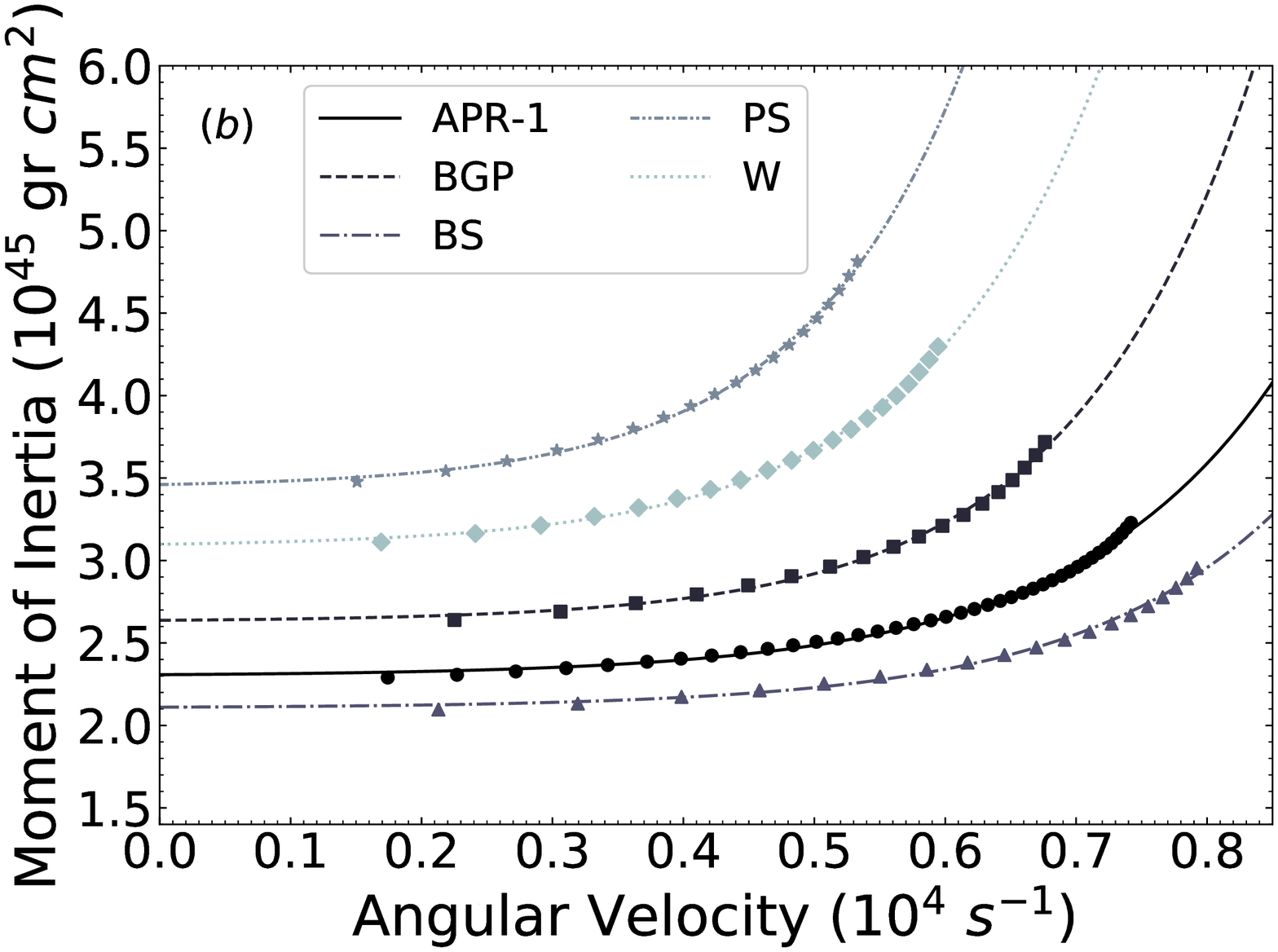}
	~
	\includegraphics[width=0.32\textwidth]{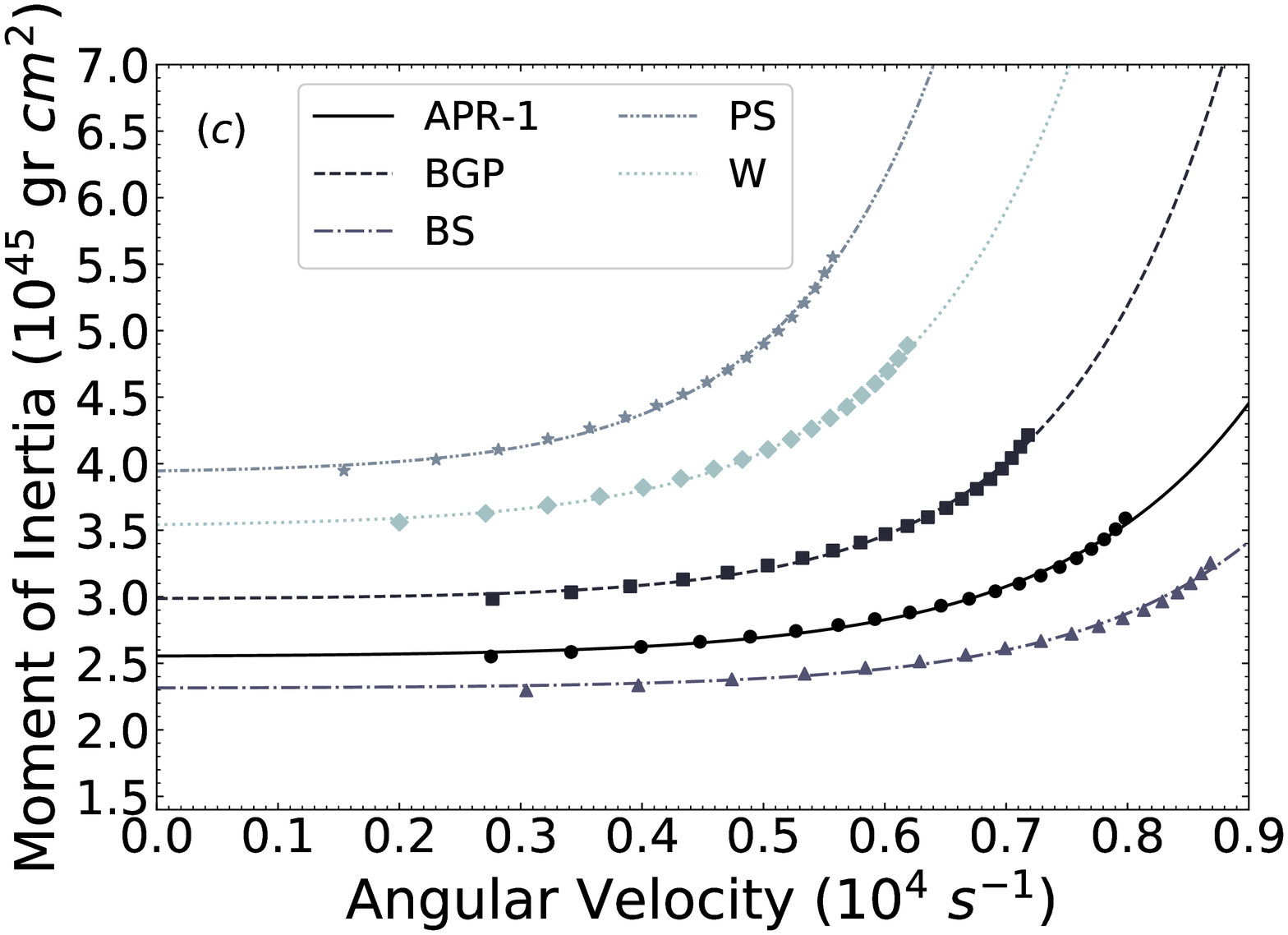}
	\caption{(color online) Constant rest mass sequences as the dependence of moment of inertia on the angular velocity  for five representative EoSs and with rest mass corresponding to (a) $M_{\rm max}^{\rm gr}=1.45 M_{\odot}$, (b) $M_{\rm max}^{\rm gr}=2 M_{\odot}$ and (c) $M_{\rm max}^{\rm gr}=2.2 M_{\odot}$. The data and fits for each EoS are presented with the circles and the solid curve for the APR-1, 
the squares and the dashed curve for the BGP, the triangles and the dashed-dotted curve for the BS, the stars and the dashed-dotted-dotted curve for the PS and the diamonds and the dotted curve for W.}
	\label{fig:bi_moi}
\end{figure*}

\subsection{Equation of state effects on the braking index of pulsars}\label{secsub:3_6}
It is well-known that the angular velocity $\Omega$ of an isolated pulsar decreases very slowly with the time. Various energy loss mechanisms are responsible for this effect, including mainly dipole radiation, charged particles ejections and gravitational waves radiation~\cite{Glendenning-2000, Weber-99,Lorimer-05,Hamil-2015,Becker-09,Lyne-015,Lorimer-2008,Manchester-2005}. In this case, and in the most simple model, the evolution of the angular velocity is given by the power law
\begin{equation}
\dot{\Omega}\equiv \frac{d \Omega}{d t}=-{\cal J}\Omega^{n}
\label{eq:brak-1}
\end{equation}

The braking index, $n$, of a pulsar, which describes the dependence of the braking torque on the rotation frequency, is a fundamental parameter of pulsar electrodynamics. Simple theoretical arguments, based on the assumption of a constant dipolar magnetic field, predict $n = 3$. It is easy to show that Eq.~\eqref{eq:brak-1} leads to the fundamental relation
\begin{equation}
n(\Omega)=\frac{\Omega\ddot{\Omega}}{\dot{\Omega}^2}=3-\frac{3\Omega I'+\Omega^2 I''}{2I+\Omega I'}
\label{eq:brak-2}
\end{equation}

\noindent where dot corresponds to the derivative with time, $I'=d I/d\Omega$ and $I''=d^2 I/d\Omega^2$. Now, considering the simple power law dependence $I \sim \Omega^{\lambda}$, the braking index takes the simple and transparent value
\begin{equation}
n(\Omega)=3-\lambda
\label{eq:brak-4}
\end{equation}

\begin{figure}[H]
	\includegraphics[width=0.5\textwidth]{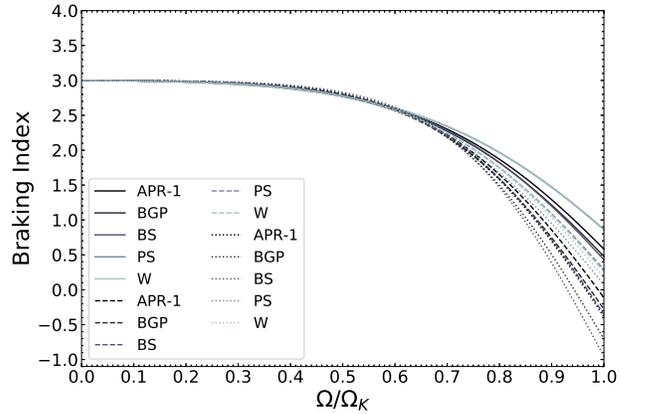}
	\caption{(color online) Braking index dependence on the angular velocity for the five representative EoSs (APR-1, BGP, BS, PS and W) with constant rest masses. The solid curves correspond to the $M_{\rm max}^{\rm gr} = 1.45M_{\odot}$, the dashed curves to the $M_{\rm max}^{\rm gr} = 2M_{\odot}$ and the dotted curves to the $M_{\rm max}^{\rm gr} = 2.2M_{\odot}$. }
	\label{fig:bi_all}
\end{figure}

While for $\lambda=0$ (moment of inertia independent from angular velocity) we recover the well-known result $n=3$, in general we expect that the inequality $n(\Omega) \leq 3$ must hold. There is a special case where for some reasons when the denominator of Eq. \eqref{eq:brak-2} goes to zero, then the braking index exhibits a singularity which leads to increasing of $\Omega$ with time~\cite{Glendenning-2000,Weber-99,Glendenning-97,Heiselberg-1998,
Zdunik-2006,Bagchi-2015}. This is an interesting effect (which may be caused due to a phase transition in the interior of a pulsar) but we are not going to study it further in this work. Instead, we studied the effects  of the EoS on the braking index as well as on the evolution of the angular velocity of a pulsar, especially for very young, at their birth, with their angular velocity being at the mass-shedding limit.\\
\indent In particular, we studied the moment of inertia dependence on the angular velocity for five representative EoSs and for three different values of rest mass. In each case, we produced a fit as shown in Fig.~\ref{fig:bi_moi}, according to the formula
\begin{equation}
I = {\rm g}_{1} + {\rm g}_{2} \exp\left[{\rm g}_{3} \Omega \right]
\label{eq:bi_moi}
\end{equation}
\noindent where $\rm g_{1}$, $\rm g_{2}$ are in units of moment of inertia $(10^{45}\rm gr \text{ } cm^{2})$ and $\rm g_{3}$ in units of time $(\rm s)$.\\
\indent In order to see how the rest mass effects the braking index, we present at Fig.~\ref{fig:bi_all} the five representative EoSs for the different rest masses.

From Fig.~\ref{fig:bi_all}, it is clear that the rest mass plays an important role on the braking index, i.e. by increasing the rest mass value, the braking index decreases more sharply. This effect will remain valid for all EoSs studied in this paper.

\section{Discussion and Conclusions}\label{sec:4}
Different sequences of uniformly rotating neutron stars have been constructed for a large number of hadronic EoSs based on various theoretical nuclear models. In this paper we have studied the bulk properties of neutron stars in correlation with the mass-shedding limit (Keplerian frequency). To be more specific, we have calculated their gravitational and rest mass, equatorial and polar radii, dimensionless angular momentum, angular velocity, moment of inertia and eccentricity. Relations between the Keplerian frequency and the bulk properties of neutron stars have been found and shown in the corresponding figures. These universal relations may help to impose constraints on the radius of a neutron star when its mass and Keplerian frequency is well fixed simultaneously. For example, this is the  case of a millisecond pulsar (in a binary system) which acquired angular momentum by accretion and becoming a maximally-rotating one with measured mass~\cite{Lattimer-2017}.\\
\indent The dependence of moment of inertia, eccentricity and Kerr parameter on the total gravitational mass at the Keplerian sequence, is also obtained. In all cases, the EoSs presented similar behavior, so as a follow up, we have studied the dependence of these parameters on the gravitational mass at the maximum mass configuration. We have concluded with this way that moment of inertia and Kerr parameter can provide us with universal relations as a function of the gravitational mass at the maximum mass configuration for the Keplerian frequency. It is also interesting the effect of the eccentricity at the maximum mass configuration for the Keplerian frequency on the corresponding gravitational mass, where it seems that eccentricity behaves as an EoS-independent property. Moreover, we have found that the Kerr parameter reaches a maximum value at around 0.75 (stiffest EoS) for neutron stars. The importance of this result falls under the fact that the gravitational collapse of a uniformly rotating neutron star, constrained to mass-energy and angular momentum conservation, cannot lead to a naked singularity, or in other words, a maximally rotating Kerr black hole~\cite{Lo-2011}.\\
\indent As a limiting case in our study, we presented also an EoS suitable to describe quark stars and one with appearance of hyperons at high densities. In non-rotating case the results are in good agreement with the hadronic EoSs where in maximally-rotating one, the difference from linearity is obvious. Moreover, as far as concerning the Kerr parameter, it is undeniable that its study on quark stars requires a different approach. Although that a thorough study is needed, we can see that the values of Kerr parameter of quark stars are significantly larger, not only of neutron stars, but also of black holes. The latter can be useful indicator to identify maximally-rotating quark stars~\cite{Lo-2011}. However, a detail study must be done in order to acquire the possible effects of the EoS on quark stars and their similarities with neutron stars. Concerning the hyperonic EoS, while its consistent with our 23 hadronic EoSs allows it to be studied with them, a detail study mainly based on hyperonic EoSs, would be more suitable.\\
\indent Normal and supramassive sequences of constant rest mass for a specific EoS have been constructed. In the corresponding figures, we present the stability and instability region of a neutron star. This is possible by plotting the evolution of a neutron star along the constant rest mass sequences. The extraordinary effect of supramassive ones, is that they can inform us for the gravitational collapse to a black hole. The gravitational collapse of a rotating neutron star to a black hole, creates a black hole with almost the same mass and angular momentum as the initial star (small amount of total mass and angular momentum carried away by gravitation radiation~\cite{Baiotti-2005}), and therefore, the same Kerr parameter. Henceforth, this effect may provide us an observable precursor to gravitational collapse to a black hole. It is important to add here that this effect will remain valid for all the EoSs studied in this paper.\\
\indent In order to imply possible constraints on the EoS, we have constructed the LSRMS for the variety of the EoSs and the dimensionless moment of inertia. In particular, we have presented them in a figure of the angular velocity as a function of the Kerr parameter and the dimensionless moment of inertia as a function of the compactness parameter, respectively. In both cases, we have extracted a window where these properties can lie. In the first case, concerning the LSRMS, because this sequence is the one that corresponds to the maximum mass configuration at the non-rotating model, this is also the window where the EoS can lie, constraining with this way, simultaneously, the angular velocity and spin parameter (or Kerr parameter) on neutron stars. In the second case, the window that is formed can help us to constrain moment of inertia and compacteness parameter. The latter, can impose strong constraints in radius of neutron stars, which is one of the open problems in nuclear astrophysics.\\
\indent Afterwards, we have updated the work of Lattimer and Prakash~\cite{Lattimer-05} by using EoSs which are in consistent with the current observed limits of neutron star mass~\cite{Demorest-010,Fonseca-016,Arzoumanian-2018,Antoniadis-013,Cromartie-19,Linares-18}. In this work we propose the possible existence of an empirical solution, similar to the Tolman VII analytical solution, for neutron stars, using as a guide the maximally-rotating configuration in order to describe both the non-rotating and the maximally-rotating configuration. The existence of this solution can help to define the ultimate density of cold baryonic matter by setting an absolute upper limit at the central energy density. The latter can be a useful insight because it can inform us on the appearance of a phase transition in the interior of the star and its leading to the back-bending phenomenon in pulsars.\\
\indent Finally, we have studied the effects of the EoSs on the braking index of pulsars. Braking index, as an intrinsic property of neutron star's structure, can inform us about the rate of change of angular velocity. Although we know it is very slow, after the 70\% of Keplerian angular velocity, braking index is undergoing significant changes through the influence of the rest mass. This specific area, from 70\% through the 100\% of Keplerian angular velocity, may provide us with useful insights on the constitution of the dense nuclear matter.\\
\indent In near future, neutron star mergers and measurements of gravitational waves, besides the fact that are a powerful tool to study compact objects, such as neutron stars and black holes, they will be able to provide us with the Keplerian frequency of these objects. In fact, the remnant formed in the immediate aftermath of the GW170817 merger although is believed to have been differentially rotating and not uniformly, it contains sufficient angular momentum to be near its mass-shedding limit~\cite{Lattimer-2019}. The observational measurement of Keplerian frequency, along with the theoretical predictions, would provide us with strong constraints on the high density part of the EoS.

\section*{Acknowledgments}
The authors would like to thank Professor N. Stergioulas for providing his code RNS and his very useful comments and clarifications. One of the authors (Ch.C.M) would like to thank the Theoretical Astrophysics Department of the University of Tuebingen, where part of this work was performed and Professor K. Kokkotas for his useful comments on the preparation of the manuscript. We also thank Dr. T. Athanasiadis for his useful corresponding. Furthermore, we would like to thank the anonymous referees for their valuable comments and suggestions which helped us to improve the present paper. \\
\indent This work was partially supported by the COST action PHAROS (CA16214) and the DAAD Germany-Greece grant ID 57340132.

\appendix
\section{The MDI model}\label{app:1}
In order to study specific properties and evolutionary process of neutron stars we employed the MDI model. The energy per particle, according to MDI, is given by~\cite{Prakash-1997,Moustakidis-15}

\begin{widetext}
	\begin{eqnarray}
	\label{e-T0}
	E_b(n,I)&=&\frac{3}{10}E_F^0u^{2/3}\left[(1+I)^{5/3}+(1-I)^{5/3}\right]+
	\frac{1}{3}A\left[\frac{3}{2}-X_{0}I^2\right]u
	+
	\frac{\frac{2}{3}B\left[\frac{3}{2}-X_{3}I^2\right]u^{\sigma}}
	{1+\frac{2}{3}B'\left[\frac{3}{2}-X_{3}I^2\right]u^{\sigma-1}}
	\nonumber \\
	&+&\frac{3}{2}\sum_{i=1,2}\left[C_i+\frac{C_i-8Z_i}{5}I\right]\left(\frac{\Lambda_i}{k_F^0}\right)^3
	\left(\frac{\left((1+I)u\right)^{1/3}}{\frac{\Lambda_i}{k_F^0}}-
	\tan^{-1} \frac{\left((1+
		I)u\right)^{1/3}}{\frac{\Lambda_i}{k_F^0}}\right)\nonumber \\
	&+&
	\frac{3}{2}\sum_{i=1,2}\left[C_i-\frac{C_i-8Z_i}{5}I\right]\left(\frac{\Lambda_i}{k_F^0}\right)^3
	\left(\frac{\left((1-I)u\right)^{1/3}}{\frac{\Lambda_i}{k_F^0}}-
	\tan^{-1}
	\frac{\left((1-I)u\right)^{1/3}}{\frac{\Lambda_i}{k_F^0}}\right)
	\end{eqnarray}
\end{widetext}

\noindent where $I=(n_{n}-n_{p})/n$, $X_{0} = x_{0} + 1/2$ and $X_{3} = x_{3} + 1/2$.\\
\indent In Eq. \eqref{e-T0}, the ratio $u$ is defined as $u=n/n_s$, with $n_s$ denoting the equilibrium symmetric nuclear matter density (or saturation density) and equals to 0.16 fm$^{-3}$. The parameters $A$, $B$, $\sigma$, $C_1$, $C_2$ and $B'$, which are called coupling constants and appear in the description of symmetric nuclear matter (SNM), are determined so that the relation $E_{b}(n_{s},0)=-16$ {\rm MeV} holds. The finite range parameters are $\Lambda_1=1.5 k_F^{0}$ and $\Lambda_2=3 k_F^{0}$ with $k_F^0$ being the Fermi momentum at the saturation density $n_s$. By suitably choosing the rest parameters $x_0$, $x_3$, $Z_1$, and $Z_2$, which appear in the description for asymmetric nuclear matter (ANM), it is possible to obtain different forms for the density dependence of symmetry energy as well as the value of slope parameter L and the value of symmetry energy at the saturation density~\cite{Prakash-1997,Moustakidis-15}. Actually, for each value of L, the density dependence of symmetry energy is adjusted so that the energy of pure neutron matter is comparable with those of the existing \emph{state-of-the-art} calculations~\cite{Prakash-1997,Moustakidis-15}.

\section{Observed frequency limit}\label{app:2}
\indent Lattimer and Prakash derived a relation in Ref.~\cite{Lattimer-2004}, which gives the Keplerian frequency of a rotating neutron star, in terms of radius $R$ and mass $M$ of the corresponding non-rotating neutron star. The relation is

\begin{equation}
f_{k} = 1045 \left(\frac{M}{M_{\odot}}\right)^{1/2} \left(\frac{10km}{R}\right)^{3/2} \quad (\rm Hz)
\end{equation}
which can be written as $f_{k}\approx 0.5701 f_{S}$, where $f_{S}$ is the Keplerian rate for a rigid Newtonian sphere, and it is given by the equation

\begin{equation}
f_{S} = 1833 \left(\frac{M}{M_{\odot}}\right)^{1/2} \left(\frac{10km}{R}\right)^{3/2} {\rm (Hz)}
\end{equation}

Following the work of Riahi {\it et al.}~\cite{Riahi-019}, in order to find a more accurate relation, we have constructed a relation, based on a three order polynomial fit in terms of mass and radius of the corresponding non-rotating neutron star, given by the form

\begin{widetext}
\begin{equation}
f_{k}/f_{S} = 0.559 + 2.69 \left(\frac{M}{M_{\odot}}\right) \left(\frac{km}{R}\right) - 20.28 \left[ \left(\frac{M}{M_{\odot}}\right) \left(\frac{km}{R}\right) \right]^{2} + 55.74 \left[ \left(\frac{M}{M_{\odot}}\right) \left(\frac{km}{R}\right) \right]^{3}
\label{eq:appb_3}
\end{equation}
\end{widetext}
with error up to 4\%, in comparison with Lattimer and Prakash where the error was up to 30\%.\\
\indent For the observed frequency of the fastest known pulsar, PSR J1748-2446ad, which rotates with a frequency of 716 Hz, we obtained the relation~\ref{eq:appb_3} and its schematic presentation is presented in Fig.~\ref{fig:mass_radius}.

\section{Analytical solution - Tolman VII}\label{app:3}
The basic ingredients of the analytical solution - Tolman VII of Einstein's equations for a non-rotating spherical symmetric object, which in this case is neutron star, are presented below.\\
\indent The metric functions are defined as follows
\begin{equation}
e^{-\lambda}=1-\beta x^2(5-3x^2), \quad e^{\nu}=\left(1-\frac{5\beta}{3}\right)\cos^2\phi,
\label{metric-el-Tolm}
\end{equation}
where
\[x = \frac{r}{R}, \quad \phi=\frac{w_1-w}{2}+\phi_1, \quad \phi_1=\tan^{-1}\sqrt{\frac{\beta}{3(1-2\beta)}}\]
 and
 \[w=\ln\left(x^2-\frac{5}{6}+\sqrt{\frac{e^{-\lambda}}{3\beta}}\right), \quad w_1=\ln\left(\frac{1}{6}+\sqrt{\frac{1-2\beta}{3\beta}}  \right).  \]
The energy density and the pressure read as
\begin{equation}
\frac{{\cal E}(x)}{{\cal E}_c}=(1-x^2), \quad {\cal E}_c=\frac{15Mc^2}{8\pi R^3},
\label{Tolm-E}
\end{equation}
\begin{equation}
\frac{P(x)}{{\cal E}_c}=\frac{2}{15}\sqrt{\frac{3e^{-\lambda}}{\beta }}\tan\phi-\frac{1}{3}+\frac{x^2}{5}.
\label{Tolm-Pr}
\end{equation}
There are some constraints related with the validity of the Tolman VII - analytical solution. In particular, the central value of pressure becomes infinite for $\beta=0.3862$, while the speed of sound remains less than the speed of light only for $\beta<0.2698$~\cite{Moustakidis-2017}. This solution leads to a stable configurations only for  $\beta< 0.3428$~\cite{Moustakidis-2017}.


\end{document}